\documentclass[]{aa} 
\usepackage{graphicx, natbib, latexsym,amssymb,amsmath} 
\usepackage{txfonts} 
\usepackage{graphicx, natbib, latexsym,amssymb} 
\bibpunct{(}{)}{;}{a}{,}{,} 
\usepackage{supertabular} 
\usepackage{longtable,lscape} 
\def\aap{A\&A}                
\def\apjl{ApJ}                
\def\apjs{ApJS}               

\def\hh13cop{$\mathrm{H^{13}CO^+}$} 
\def\n2hp{$\mathrm{N_2H^+}$} 
\def\c18o{$\mathrm{C^{18}O}$} 
 
\def\h2co{$\mathrm{H_2CO}$}

\def\degr{$^\circ$}

\begin{document} 
   \title{Parallaxes and proper motions of interstellar masers toward the Cygnus~X star-forming complex}
   \subtitle{I. Membership of the Cygnus~X region}                                      
   \author{K.~L.~J.~Rygl \inst{1, 2}, 
           A.~Brunthaler\inst{2, 3}, 
           A.~Sanna\inst{2}, 
           K.~M.~Menten\inst{2}, 
           M.~J.~Reid\inst{4}, 
           H.~J.~van Langevelde\inst{5,6}, 
           M.~Honma\inst{7}, 
           K.~J.~E.~Torstensson\inst{6,5}, 
           \and K.~Fujisawa\inst{8} 
                    } 
 
   \institute{Istituto di Fisica dello Spazio Interplanetario (INAF-IFSI),
              Via del fosso del cavaliere 100, 00133 Roma, Italy\\ 
              \email {kazi.rygl@inaf.it} 
             \and Max-Planck-Institut f\"ur Radioastronomie (MPIfR), 
              Auf dem H\"ugel 69, 53121 Bonn, Germany\\ 
              \email{[brunthal,asanna,kmenten]@mpifr-bonn.mpg.de} 
              \and National Radio Astronomy Observatory, 1003 Lopezville Road, Socorro 87801, USA
              \and Harvard Smithsonian Center for Astrophysics, 60
                   Garden Street, Cambridge, MA 02138, USA\\                               
              \email{reid@cfa.harvard.edu} 
              \and Joint Institute for VLBI in Europe, Postbus 2,
                   7990 AA Dwingeloo, the Netherlands\\                                    
              \email{langevelde@jive.nl} 
              \and Sterrewacht Leiden, Leiden University, Postbus
                   9513, 2300 RA Leiden, the Netherlands\\                                 
              \email{kalle@strw.leidenuniv.nl} 
              \and Mizusawa VLBI Observatory, National Astronomical
                  Observatory of Japan, 2-21-1 Osawa, Mitaka, Tokyo 181-8588, Japan\\    
              \email{mareki.honma@nao.ac.jp} 
              \and Faculty of Science, Yamaguchi University, 1677-1
                   Yoshida,Yamaguchi 753-8512, Japan\\                                     
             \email{kenta@yamaguchi-u.ac.jp} 
             } 
 
   \date{Received ; accepted} 
  \abstract 
    {Whether the Cygnus~X complex consists of one physically connected region of star formation or of multiple independent regions projected close together on the sky has been debated for decades. The main reason for this puzzling scenario is the lack of trustworthy distance measurements.}                                             
   {We aim to understand the structure and dynamics of the star-forming regions
toward Cygnus~X by accurate distance and proper motion measurements.}   
   {To measure trigonometric parallaxes, we observed 
6.7\,GHz methanol and 22\,GHz water masers
with the European VLBI Network and the Very Long Baseline Array.}                   
   {We measured the trigonometric parallaxes and proper motions of
five massive star-forming regions toward the Cygnus~X complex and
report the following distances within a 10\% accuracy: $1.30^{+0.07}_{-0.07}$\,kpc for W\,75N,
$1.46^{+0.09}_{-0.08}$\,kpc for DR\,20, $1.50^{+0.08}_{-0.07}$\,kpc
for DR\,21, $1.36^{+0.12}_{-0.11}$\,kpc for IRAS\,20290+4052, and
$3.33^{+0.11}_{-0.11}$\,kpc for AFGL\,2591. While the distances of W\,75N, DR\,20, DR\,21, and IRAS 20290+4052 are consistent with a single distance of $1.40\pm0.08$\,kpc for the Cygnus~X complex, AFGL\,2591 is located at a much greater distance than previously assumed.
The space velocities of the four star-forming regions in the Cygnus~X complex do not suggest 
an expanding Str\"omgren sphere.} 
   {} 
 
 \authorrunning{K.~L.~J.~Rygl et al.} 
 \titlerunning{Parallax and proper motions of masers toward the Cygnus~X complex}                                                              
   \maketitle 
%
 
\section{Introduction} 
In the early days of radio astronomy, a conspicuously strong, extended 
source of radio emission was found at Galactic longitude $\sim$$80^\circ$ 
and named the Cygnus~X region (\citealt{piddington:1952}), which also stands out in infrared surveys of the Galaxy 
(\citealt{odenwald:1993}; see e.g., the 
spectacular Spitzer imaging in \citealt{kumar:2007}). 
All phases of star formation and stellar evolution are observed projected across the Cygnus~X region, including a population of dense, massive, and dusty cores with embedded protoclusters and high-mass protostellar objects (\citealt{sridharan:2002,beuther:2002,motte:2007}), ultracompact HII regions (\citealt{downes:1966, wendker:1991,cyganowski:2003}), hundreds of OB-type stars (of which $\sim$65 O-type;  \citealt{wright:2010} and references therein),
and some supernova remnants (\citealt{uyaniker:2001}).                           

The proximity of a large number of OB associations and molecular cloud complexes on the sky
motivated the explanation of Cygnus~X  as consisting of various objects at different distances
seen superposed on one another (e.g., \citealt{dickel:1969}, \citealt{pipenbrink:1988}, and \citealt{uyaniker:2001}).
Recently, the CO imaging survey of \citet{schneider:2006} has rekindled the idea of Cygnus~X as one large star-forming complex, which was already suggested in the sixties by, e.g.,  \citet{veron:1965}.
Obviously, both scenarios depend strongly on the distances measured to the individual parts of the Cygnus~X complex, because sources at a galactic longitude of $\sim80^\circ$ could
be in the Local Arm (also named the Orion or Local Spur) and nearby ($\sim$1$-$2 kpc), in the Perseus Arm at
$\sim$5 kpc, or even in the Outer Arm at distances of $\sim$10 kpc (e.g., the 
Cygnus~X-3 microquasar).

Unfortunately, distances to the Cygnus~X objects are very difficult to obtain and have large uncertainties. First, the Cygnus~X OB associations are too far for a parallax measurement with  the {\it Hipparcos} satellite, which measured distances of nearby OB associations out to a distance of 650\,pc (\citealt{zeeuw:1999}).
Second, at the Galactic longitude of Cygnus~X, the radial velocity difference between the Sun and the Cygnus~X region is close to the typical velocity
dispersion of interstellar gas in a high-mass star-forming region (SFR) (1--7\,$\mathrm{km~s^{-1}}$, \citealt{moscadelli:2002}) for distances up to 4\,kpc. Therefore, kinematic distances, which depend on the radial velocity, are not reliable for distances below 4\,kpc toward this longitude, and most distance measurements rely on the spectroscopy and photometry of
stars. 
But, these estimates are also affected by large uncertainties ($>30$\%) because the extinction toward Cygnus~X is very high and variable (\citealt{schneider:2006}).  While one can find distance estimates between
1.2 and 2\,kpc in the literature, the generally adopted value is 1.7\,kpc,
based on the spectroscopy and photometry of the stars in the
Cyg OB2 association by \citet{torres:1991} and \citet{massey:1991}. More
recently, a nearer distance of 1.5\,kpc was  obtained by \citet{hanson:2003}
using new MK stellar classification spectra of the stars in Cyg OB\,2. 
Of course, using a Cyg OB\,2 distance for the entire complex
assumes that Cygnus~X is a physically connected SFR.

The distance to this ``mini-starburst'' region is crucial for the many star formation studies performed toward it for its richness and short distance from the Sun.
As said, the distances are very uncertain: for example, three OB associations (Cyg OB\,1, 8, and 9) have distance estimates between
1.2 and 1.7 kpc, a difference of more than 30\%. This distance range
of 500\,pc is almost ten times more than the extent of the Cygnus~X region
on the sky -- 4$^\circ$ by 2$^\circ$ or 100 by 50\,pc at a distance of 1.5\,kpc. 
Therefore, important physical parameters of objects in this region, such as luminosities
and masses are uncertain by a factor of $\sim$1.7 given a distance uncertainty of 30\%. For some of the SFRs, the distance estimates have a much wider range, namely from $\sim$1\,kpc (AFGL\,2591) to 3\,kpc (DR\,21).
To distinguish whether all clouds are at the same distance or are only projected close together on the plane of the sky, a direct estimate of distances to distinct objects toward Cygnus~X is required.

In this context, we used strong 6.7\,GHz methanol masers and 22\,GHz water masers as astrometric targets to measure distances of five distinct SFRs toward the Cygnus~X complex. No previous parallax measurements were carried out toward this region.
Using the terminology of \citet{schneider:2006}, the Cygnus~X complex is
divided about the Cyg OB\,2 cluster at $(l, b)$ = ($80{\rlap{$.$}\,^\circ 22}, +0{\rlap {$.$}\,^\circ}80$) into a northern region, 
at Galactic longitudes greater than about 80$^\circ$, and a southern region, at lower
longitudes. Methanol maser emission was observed toward four SFRs with the European VLBI Network (EVN): W\,75N, DR\,20, and DR\,21 in Cygnus~X North, and IRAS\,20290+4052 (which is likely part of the
Cyg OB\,2 association, see \citealt{odenwald:1989,parth:1992}, hereafter IRAS\,20290) in Cygnus~X South. The pioneering work presented in \citet{rygl:2010a} demonstrates the capability of the EVN to achieve parallax accuracies as good as 22\,$\mu$m. Water maser emission was observed toward AFGL\,2591, which is projected within Cygnus~X South, with the Very Long Baseline Array (VLBA).
 
The observations presented here comprise one year of VLBA observations
and two years of EVN observations with the addition of Japanese
antennas (the Yamaguchi 
32-m antenna and the Mizusawa station of the VLBI Exploration of Radio Astrometry, VERA) for several epochs to increase the angular resolution.  We report on the preliminary distances from the first two-year results, using only the EVN antennas, and the evidence that AFGL\,2591 is really projected against the Cygnus~X region, hence not part of it. To optimize the results with the long baselines afforded by
the Japanese antennas, we need to develop more sophisticated calibration procedures, since several maser spots resolve on the $\sim$9000\,km baselines.
This analysis will be included in a future paper once the observations are completed.

\section{Observations and data reduction} 
 
\begin{table*} 
\centering 
\caption{ Source parameters of the EVN and the VLBA (AFGL\,2591) observations\label{ta:sources}} 
\begin{tabular}{l c c r r l r c l} 
\hline\hline 
\noalign{\smallskip} 
Source  & R.A. (J2000) & Dec. (J2000) & \multicolumn{1}{c}{$l$}&\multicolumn{1}{c}{$b$}&\multicolumn{1}{c}{$\phi$ \tablefootmark{a}} &P.A.\tablefootmark{a} & Brightness\tablefootmark{b}& Restoring Beam\tablefootmark{b}\\ 
& (h:m:s) & ($^\circ$ : ' : '')& \multicolumn{1}{c}{(\degr)} & \multicolumn{1}{c}{(\degr)} &\multicolumn{1}{c}{(\degr)} & \multicolumn{1}{c}{(\degr)} & ($\mathrm{Jy~beam^{-1}}$)& (mas, mas, deg)\\                                                     
\noalign{\smallskip} 
\hline 
\noalign{\smallskip} 
W\,75N &  20:38:36.426 & 42:37:34.80 & 81.87&0.78&0 & \multicolumn{1}{c}{--}& $0.2-23.8$&4.87, 3.65,
--50.7  \\                                                       
DR\,21A & 20:39:01.993 & 42:24:59.29 &81.75 &0.59&0.2 & 159.5 & $0.1 - 1.4$&4.95,
3.80, --50.0\\                                                          
DR\,21B &  20:39:00.376 & 42:24:37.11 &81.74&0.59&0.2& 161.3 & $0.2-1.3$&4.90,
3.77, --52.1\\
DR\,21(OH) & 20:39:01.057  & 42:22:49.18 &81.72&0.57& 0.2 & 162.9 &$0.02-0.15$ &  4.90, 3.80, --51.3\\                                                               
DR\,20   &  20:37:00.962 & 41:34:55.70 &80.86&0.38& 1.1 &--164.3 & $0.04-0.8$&4.91,
3.82, --50.8\\                                                          
IRAS\,20290+4052  & 20:30:50.673 &41:02:27.54 & 79.74&0.99& 2.1 &--138.0&$0.1-2.3$
&4.91, 3.74, --48.6\\                                                  
J2045+4341 &  20:45:07.055 & 43:41:44.51 &83.44&0.50& 1.6 & 48.0& 0.0027&6.02,
4.47, --39.8\\                                                          
J2048+4310 &  20:48:19.521  & 43:10:42.09 &83.41&--0.28& 1.8 & 72.8 &0.011&4,87,
3,48, --48.0\\                                                          
J2029+4636 &  20:29:18.937 &  46:36:02.25 &84.08&4.48& 4.3 & 23.3 &0.041&4.71,
3.27, --51.4\\                                                          
\noalign{\smallskip} 
\hline 
\noalign{\smallskip} 
AFGL\,2591 & 20:29:24.823 & 40:11:19.59 &78.89&0.71& 0 &\multicolumn{1}{c}{--} &17.7 & 0.80, 0.40,
--14.9\\                                                                
J2007+4029 & 20:07:44.945 & 40:29:48.60 &76.82&4.30& 4.1&--63.0 & 0.992 & 0.79,
0.40, --18.6\\                                                          
J2033+4000 & 20:33:03.669 & 40:00:24.33 &79.15&0.04& 0.7 & 125.0 & 0.025 & 0.81, 0.40, --17.9\\                                                          
J2032+4057 & 20:32:25.771 & 40:57:27.83 & 79.85&0.70& 1.0  &36.8 &\multicolumn{1}{c}{--} &\multicolumn{1}{c}{--} \\ 
J2033+4040 & 20:32:45.342 & 40:39:38.12 &79.85&0.70& 0.8 & 53.5 & \multicolumn{1}{c}{--} & \multicolumn{1}{c}{--} \\ 
\noalign{\smallskip} 
\hline 
\end{tabular} 
\tablefoot{
\tablefoottext{a}{Separation, $\phi$, and position angles (east of north), P.A.\,, between the phase reference sources, maser W\,75 N or AFGL\,2591, and the target. }
\tablefoottext{b}{The brightness and restoring beam size (east of north) are listed for the first epoch.}}
\end{table*} 

\subsection{EVN methanol maser observations} 
The EVN\footnote{The European VLBI Network is a joint facility of European, Chinese, South African, and other radio astronomy institutes funded by their national research councils.} observations were carried out in eight epochs between March 2009 and November 2010 under project EB039.
These dates were scheduled near the minima and maxima
of the sinusoidal parallax signature in right ascension to optimize
the sensitivity of the parallax measurement. The parallax signature was followed for two years with a quasi symmetric coverage of t=0.0, 0.2, 0.63, 0.64 years with an equal sampling of the minima and maxima.   
These two conditions allow one to separate the proper motion from the parallax signature.                          
Each observation lasted 12 hours and made use of {\it geodetic-like} observing blocks to calibrate the tropospheric
zenith delays                                                           
at each antenna (see \citealt{reid:2004,brunthaler:2005,reid:2009a}
for a detailed discussion). 

The methanol masers were first selected from the \citet{pestalozzi:2005} 
database and then observed with the Expanded Very Large Array (EVLA, program AB1316) 
to obtain accurate positions (\citealt{xu:2009b}). To find extragalactic background sources to serve as 
reference positions, to verify their compact emission, and to obtain obtain position with subarcsecond accuracy, we observed compact NVSS (\citealt{condon:1998}) 
sources within 2$^\circ$ of W\,75N
at 5\,GHz with the EVN in eVLBI mode on December 4, 2008 (program EB039A). These observations revealed two 
compact background sources, J2045+4341 and J2048+4310 (hereafter J2045 and J2048); we also used J2029+4636 (herafter J2029) 
from the VLBA calibrator survey (\citealt{beasley:2002}) separated by 4$\rlap{$.$}\,^\circ 3$
from W\,75N.             
A typical EVN observing run started and ended with a $\sim$1 hour  
\emph{geodetic-like} observing block and about ten minutes for fringe-finder
observations of 3C\,454.3 and J2038+5119.  The                            
remaining time was spent on maser/background-source phase-referencing 
observations. 
The 6.7 GHz masers in Cygnus~X and the three background sources were
phase-referenced to the strongest maser in W\,75N, using a switching
cycle of 1.5 minutes.  Table \ref{ta:sources} lists the source
positions, separations from W\,75N, brightnesses,
and restoring beam information. DR\,21A, DR\,21B, and DR\,21(OH) are three masers thought to belong to the same SFR, but they were observed separately because their separation was at the border of the field of view limited by time-smearing ($\sim$37 arcseconds).                                         
 
The observations were performed in dual circular polarization and two-bit Nyquist sampling, 
for an aggregate recording rate of 512\,Mbps. The data were correlated 
in two passes at the Joint Institute for VLBI in Europe (JIVE) using
a one-second averaging time. The maser data                              
were correlated using one 8\,MHz band with 
1024 spectral channels, resulting in a channel separation of 7.81\,kHz 
and a channel width of 0.35\,$\mathrm{km~s}^{-1}$. The background 
sources were correlated in continuum mode with eight subbands of 8\,MHz bandwidth 
and a channel width of 0.25\,MHz. The data were reduced using the NRAO's 
Astronomical Image Processing System (AIPS) and ParselTongue (\citealt{kettenis:2006}).
The data were phase-referenced to the W\,75N maser at $v=7.1\,\mathrm{km~s^{-1}}$ and the solutions were transferred to the other masers and to the continuum background sources. More details on the reduction and calibration can be found in \citet{rygl:2010a}.

\subsection{VLBA water maser observations} 
 
We performed water maser observations using the National Radio Astronomy Observatory's\footnote{The National Radio Astronomy Observatory is a facility of the National Science Foundation operated under cooperative agreement by Associated Universities, Inc.} VLBA in four epochs between November 2008 and November 2009  under program BM272H.
The observation dates were chosen to match the minima and maxima
of the sinusoidal parallax signature in right ascension for optimal
parallax and proper motion analysis, as discussed in \citet{sato:2010}. 
For the VLBA observations,  the parallax sampling was symmetrically covered at t=0, 0.49,0.51, 1.0 years for one year.

To find suitably compact background sources near the maser
target, we observed a sample of unresolved NVSS sources with
the VLA in BnA configuration on October 5, 2007, at X and K band (program BM272). With
these observations, we measured source compactness and spectral index 
and selected the strongest, most compact ones to serve
as background sources (position references) for the parallax observations.
Additionally, the VLA data were used to obtain subarcsecond accurate positions
for the background sources and the masers (which were also included
in these VLA observations).  The positions, separations from the maser target, 
brightnesses, and restoring beam information are given in Table \ref{ta:sources}.
 
The VLBA observations, performed at the water maser frequency of 22.235 GHz, included four {\it geodetic-like} blocks
for calibrating the zenith delays at each antenna.  Two
fringe-finders (3C\,454.3 and 3C\,345) were observed at the beginning
and in the middle of each run.  
The water maser was observed together with four background sources,
one ICRF calibrator J2007+4029 (\citealt{ma:1998}, hereafter J2007), and three quasars selected from the
VLA observations of NVSS sources: J2033+4000 (hereafter J2033), J2032+4057, and J2033+4040.
We performed phase-referencing observations by fast-switching (every
30 seconds) between the maser and each of the four background sources.          
We used four adjacent subbands of 8\,MHz bandwidth in dual circular polarization.
Each subband was correlated to have 256 spectral channels, giving a channel width of 0.42\,$\mathrm{km~s}^{-1}$. The data were correlated with the VLBA correlation facility in Socorro
using an averaging time of 0.9 seconds. The calibration was carried
out in AIPS following the procedure described in \citet{reid:2009a}.    

\section{Parallax fitting}

\begin{figure*} 
\centering 
\includegraphics[width=8cm,angle=-90]{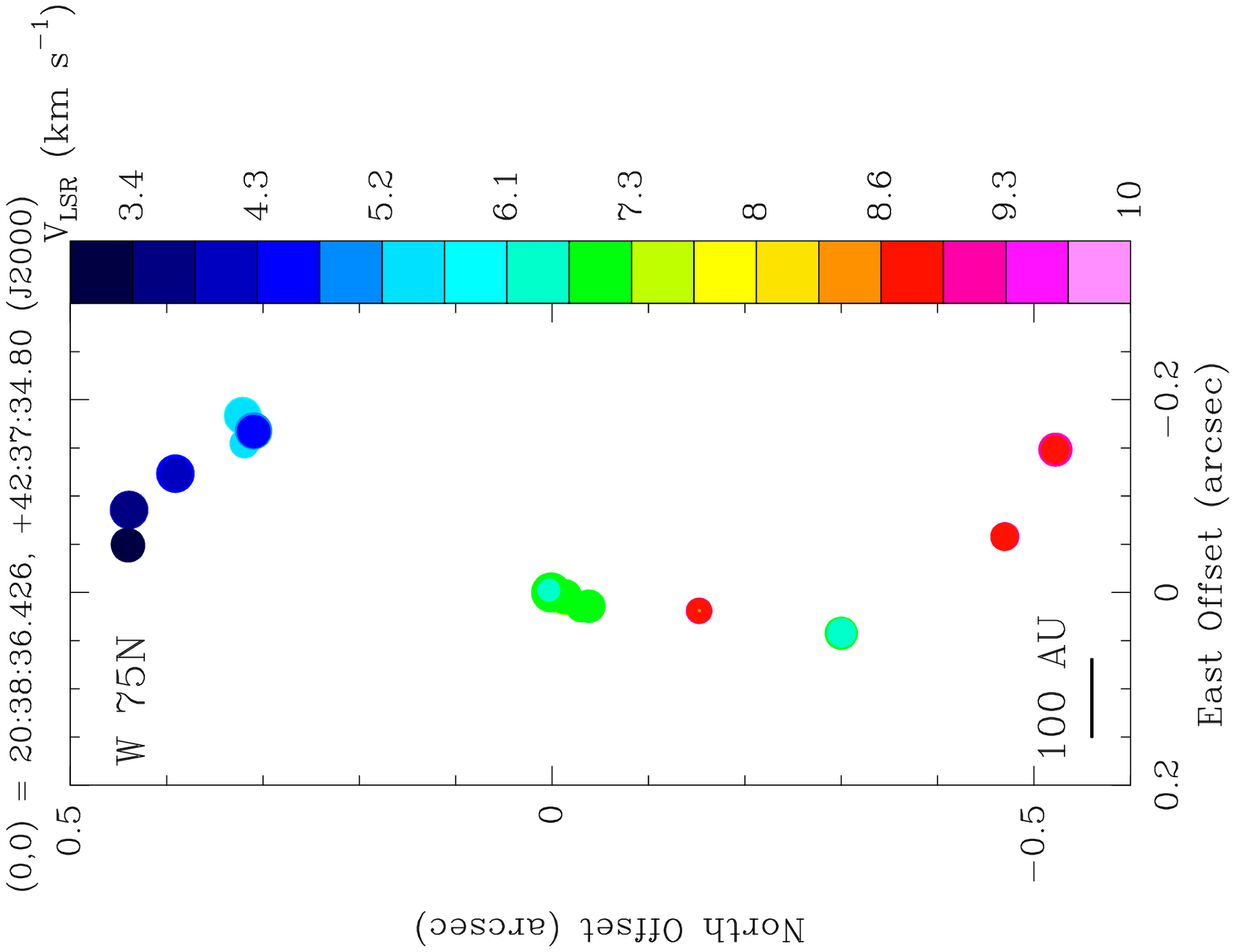} 
\includegraphics[width=8cm,angle=-90]{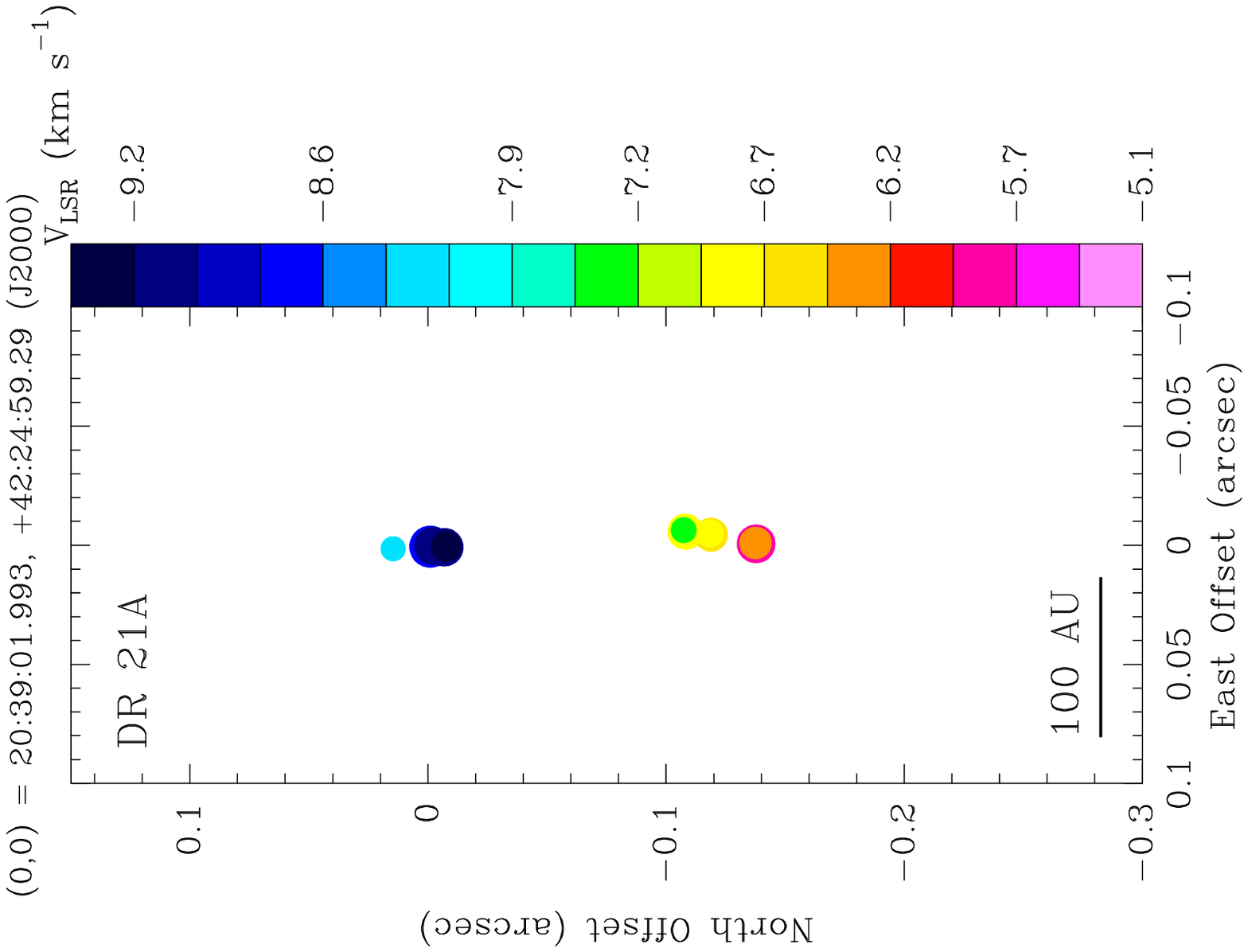} 
\includegraphics[width=8cm,angle=-90]{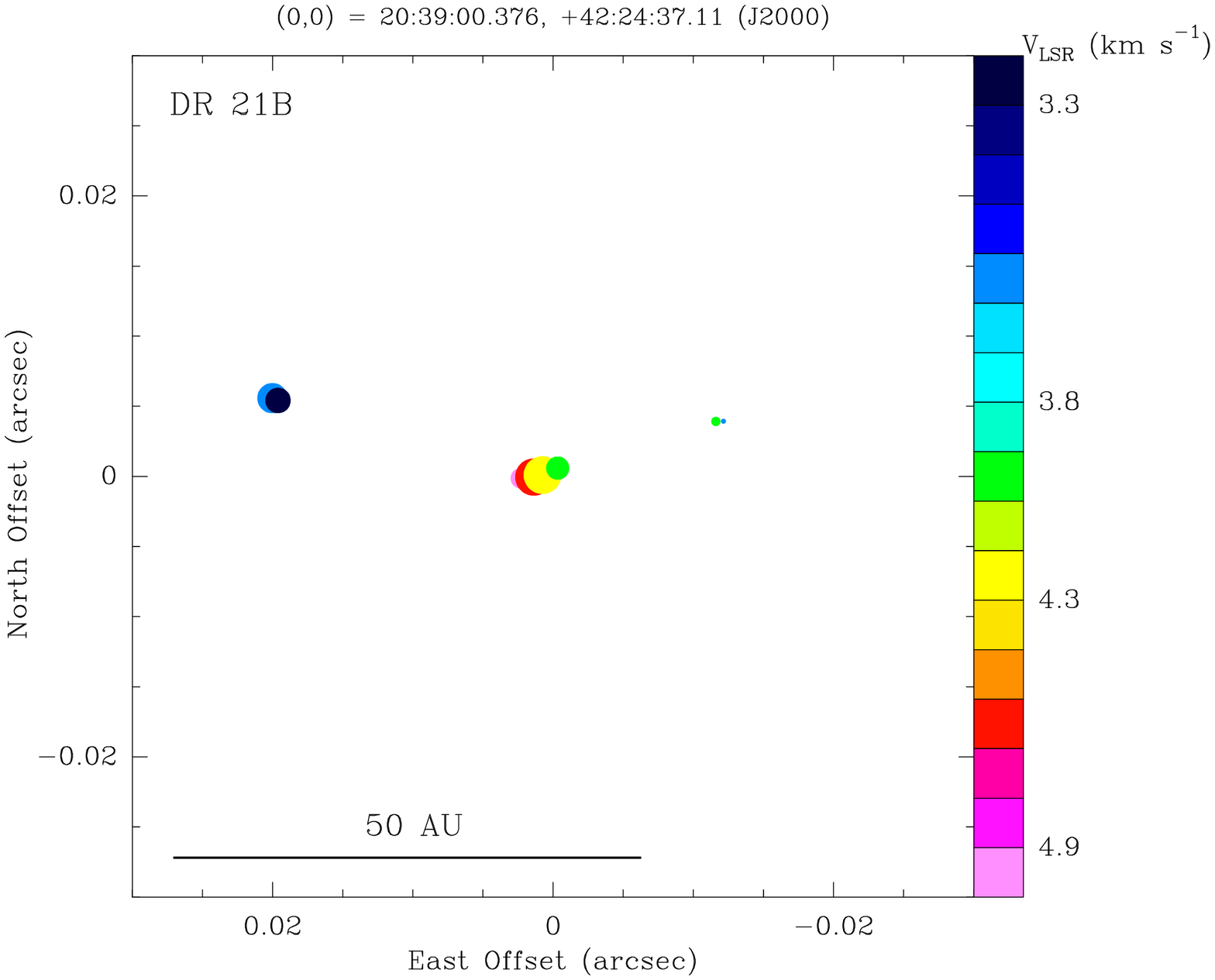} 
\includegraphics[width=8cm,angle=-90]{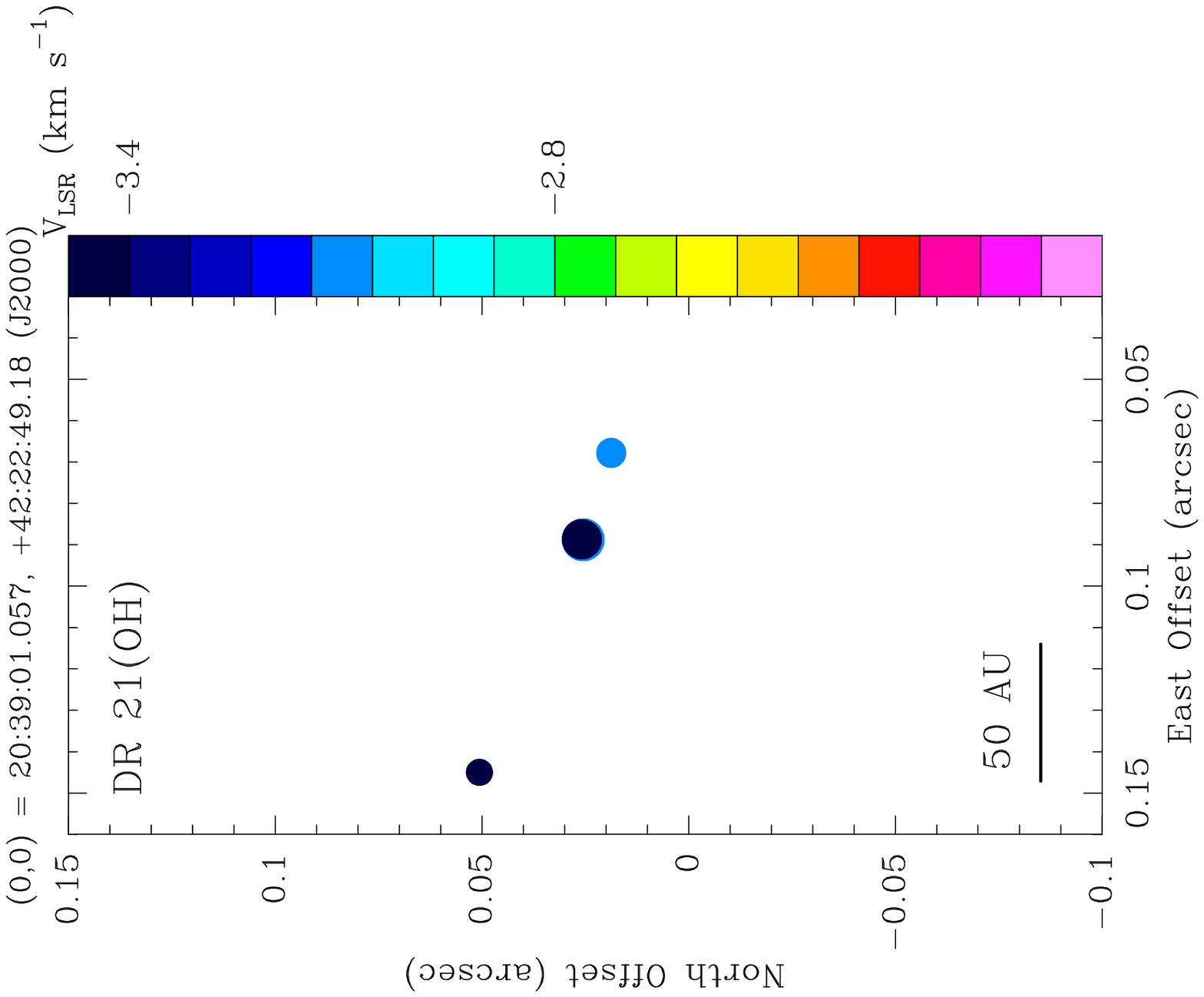} 
\includegraphics[width=8cm,angle=-90]{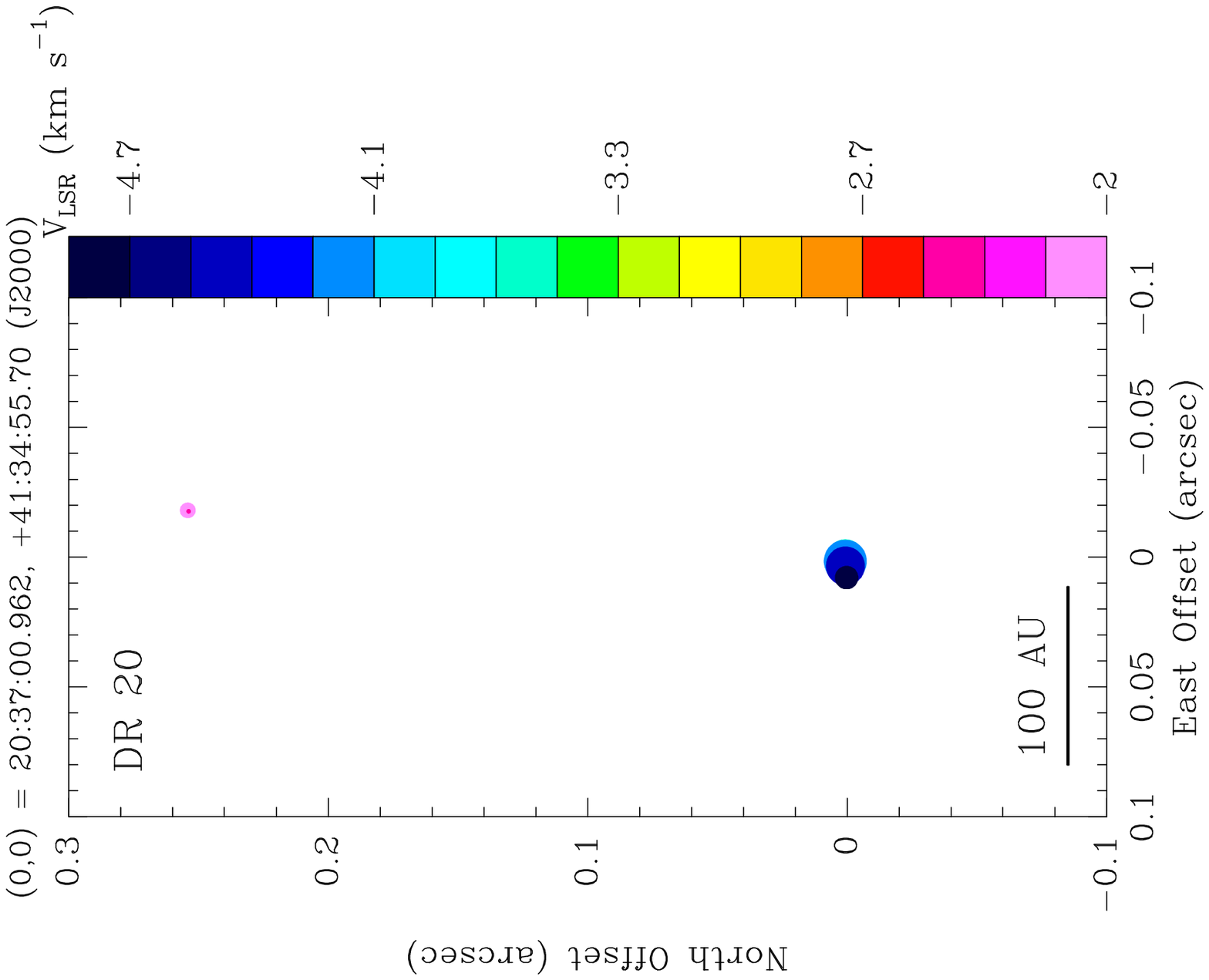} 
\includegraphics[width=8cm,angle=-90]{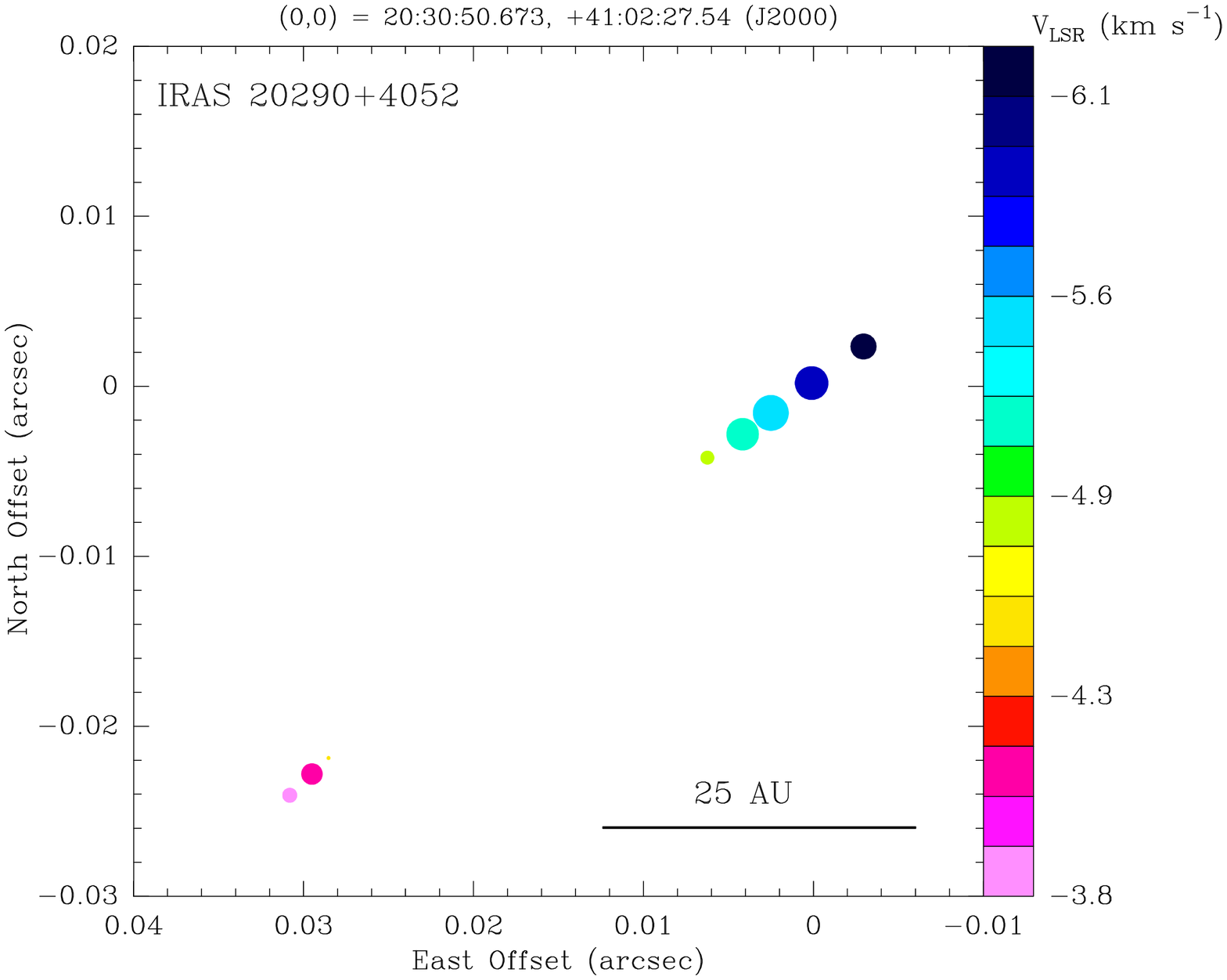} 
\caption{6.7 GHz maser spot maps for W\,75N, DR\,21A, DR\,21B, DR\,21(OH), DR\,20,
and IRAS 20290 from the first epoch of the EVN observations. The spot size is (logarithmicaly) scaled to the peak intensity of the maser spot. \label{fig:masers}}                                  
\end{figure*}

Generally, we detected 6.7 GHz methanol and 22 GHz water maser spots in several velocity 
channels each and toward multiple locations on the map (see Fig.\,\ref{fig:masers} for the methanol masers, and see \citealt{sanna:2011} for the water masers). 
All positions were determined by fitting a 2-D Gaussian brightness distribution to selected regions of the maps using the AIPS task ``JMFIT".
The parallaxes and proper motions were determined from the change
in the positions of the maser spots
relative to the background sources (position reference). 
We fitted the data with a sinusoidal parallax signature and a linear proper
motion. When the minima and maxima of the parallax signature are
sampled equally, the proper motion and parallax are uncorrelated.               
Maser spots with 
strongly nonlinear proper motions or a large scatter of position about
a linear fit were discarded, since these usually reflect spatial and spectral
blending of variable maser features and cannot be used for parallax measurements.
Only compact maser spots with well-behaved
residuals were used for the parallax fitting.

The formal position errors can underestimate the true uncertainty on the position, since they are only
based on the signal-to-noise ratios determined from the images and
do not include possible systematic errors (e.g., from residual delay    
errors). Therefore, to allow for such systematic uncertainties, 
we added error floors in quadrature to the position errors; i.e., we increased the positional error for all the epochs by a fixed amount, until reduced 
$\chi^2$ values were close to unity for each coordinate. 
The parallax fitting was then performed following the same procedure as in \citet{rygl:2010a}:
1) we performed single parallax fits per maser spot to a background
source; 2) we fitted all the maser spots (of one maser source) together in a combined parallax fit; 3) when a maser had three or more maser spots, we performed
a fit on the ``averaged data'' (see \citealt{bart:2008,hachisuka:2009}).  
However, different maser spots can be correlated because an unmodeled atmospheric delay will affect all the maser spots of one maser source in the same way. Therefore, we multiplied the uncertainty of the combined fit by $\sqrt N$, where $N$ is the number of maser spots to allow for the likelihood of highly correlated differential positions (procedure 2). 
To obtain the averaged positions for a maser spot (procedure 3), we performed parallax
fits on all the individual spots and removed their position offsets
and proper motions, after which we averaged the positions of each
epoch. The last approach has the advantage of reducing the random
errors, introduced by small variations in the internal spot distribution for a maser feature (e.g., \citealt{sanna:2010}), while leaving
the systematic errors unaffected.                                       

\begin{figure}
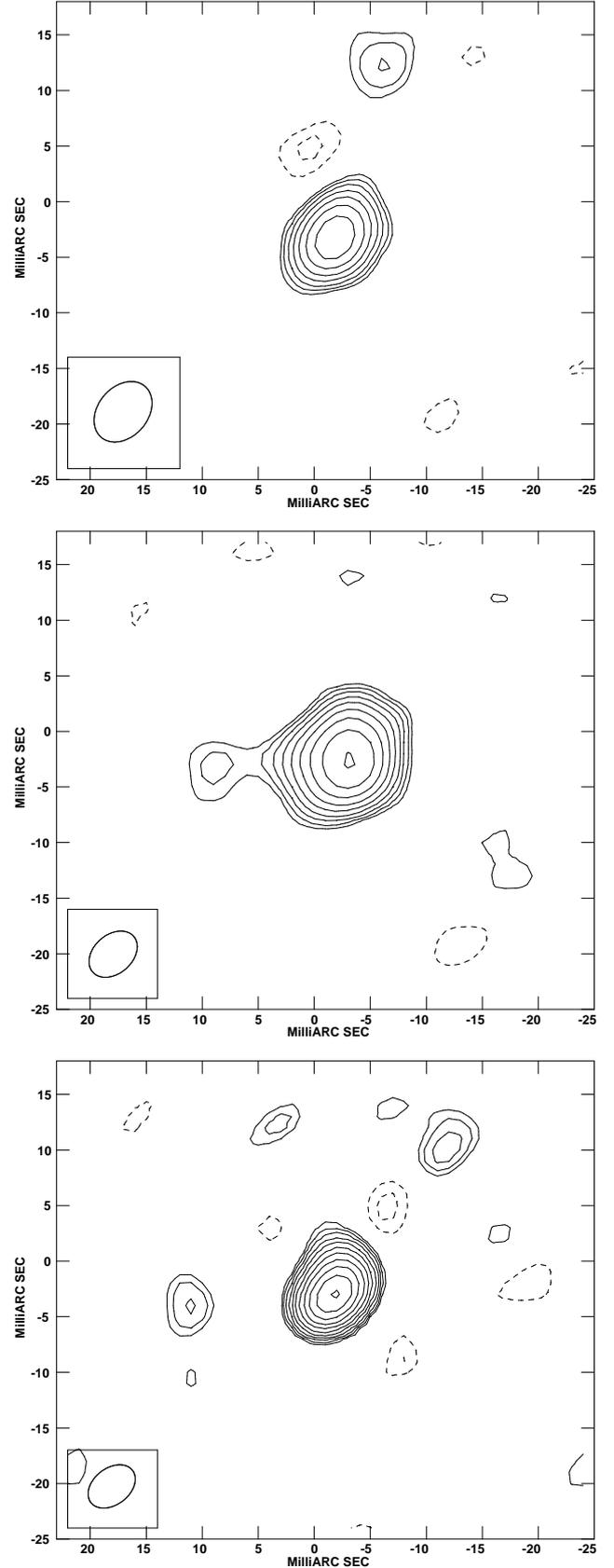
 
\centering 
\includegraphics[width=7.5cm, angle=-90]{2045_epochb.eps} 
\includegraphics[width=7.5cm, angle=-90]{2048_epochb.eps} 
\includegraphics[width=7.5cm, angle=-90]{2029_epochb.eps} 
\caption{Phase-referenced images of the background sources J2045
({\it top}), J2048 ({\it middle}), and J2029 ({\it bottom}) from the first epoch of the EVN observations. The images are in milliarcseconds offset to the positions in Table \ref{ta:sources}. The contours start at a 3$\sigma$ level, namely
$2.4\times10^{-4}$, $6.6\times10^{-4}$, and $12\times10^{-4}\,\mathrm{Jy\,beam^{-1}}$,
respectively, and increase by $\sqrt{2}$. The first negative contour ($-3\sigma$) is shown by dashed contours. The synthesized beam is shown in the bottom left corner.\label{fig:qsoima}}           
\end{figure}

\begin{figure} 
\center 
\includegraphics[width=7cm]{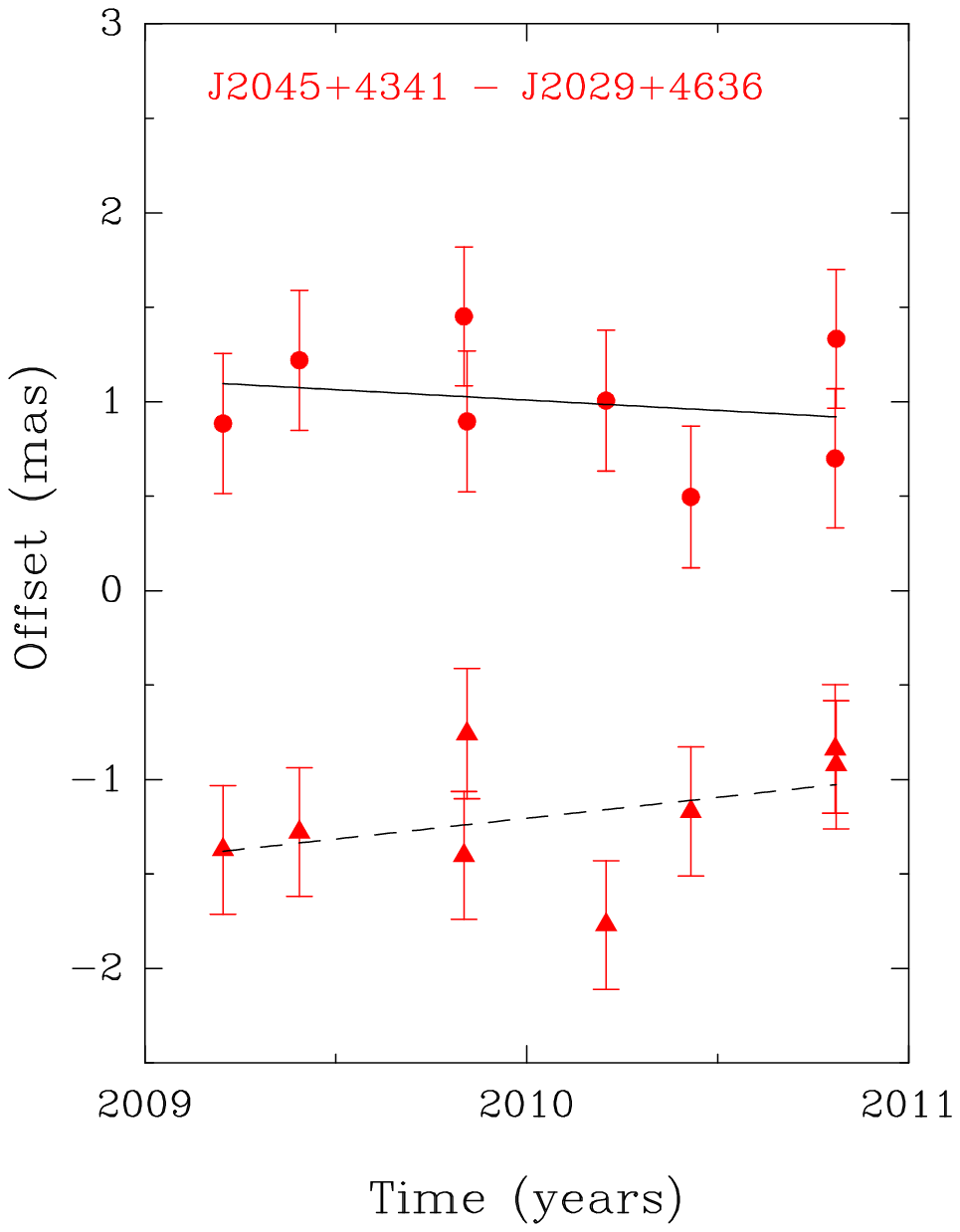} 
\includegraphics[width=7cm]{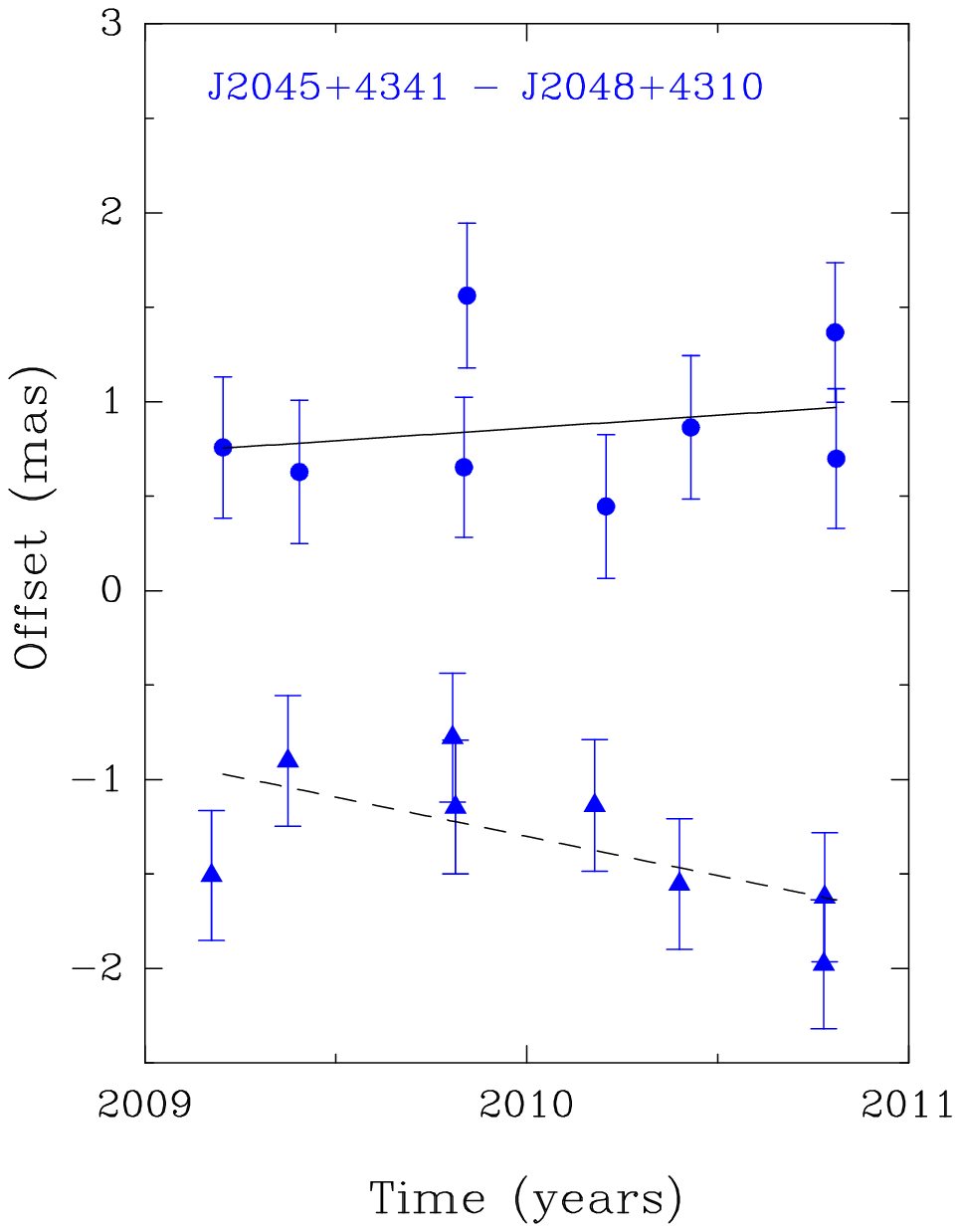} 
\caption{\label{fig:qso} Variation in time of the relative positions of background
sources J2029 ({\it top panel,} red) and J2048 ({\it bottom
panel,} blue) with respect to J2045. The dots
mark the right ascension data points, while the filled triangles
represent the declination data points. The solid line shows the right
ascension fit, while the dashed line represents the declination fit.}   
\end{figure}

The EVN observations used three background sources, of which only
one, J2045, was usable as a position reference: this quasar was
close to the maser at the phase reference position and unresolved (Fig. \ref{fig:qsoima}).
Quasar J2029 had a too large separation from W\,75N for astrometric
purposes (4$\rlap{$.$}\,^\circ$3), which can be seen from the near-milli-arcsecond scatter in
the variation of the relative position versus time of J2009 with respect
to J2045 (see Fig.\,\ref{fig:qso}). The relative position
of J2048 with respect to J2045 (also Fig.\,\ref{fig:qso}) also
shows a large scatter, thanks to the resolved emission of J2048 (Fig.\,\ref{fig:qsoima}).
Previously, we showed in \citet{rygl:2010a} that at 6.7\,GHz some quasars
can mimic an apparent motion due to changes in the quasar structure.
Two or more quasars allow one to quantify this effect so that the uncertainty in the
maser proper motion, which is the error from the derived maser proper motion fit and the apparent motion of the quasar, can be estimated. 
From the plots
of relative positions between J2045 and J2029 (J2048 was not considered because it was resolved), we estimated
the apparent relative motion to be near zero; $-0.1\pm0.2$ in right ascension and $0.2\pm0.2\,\mathrm{mas\,yr^{-1}}$ in declination.

\begin{figure}
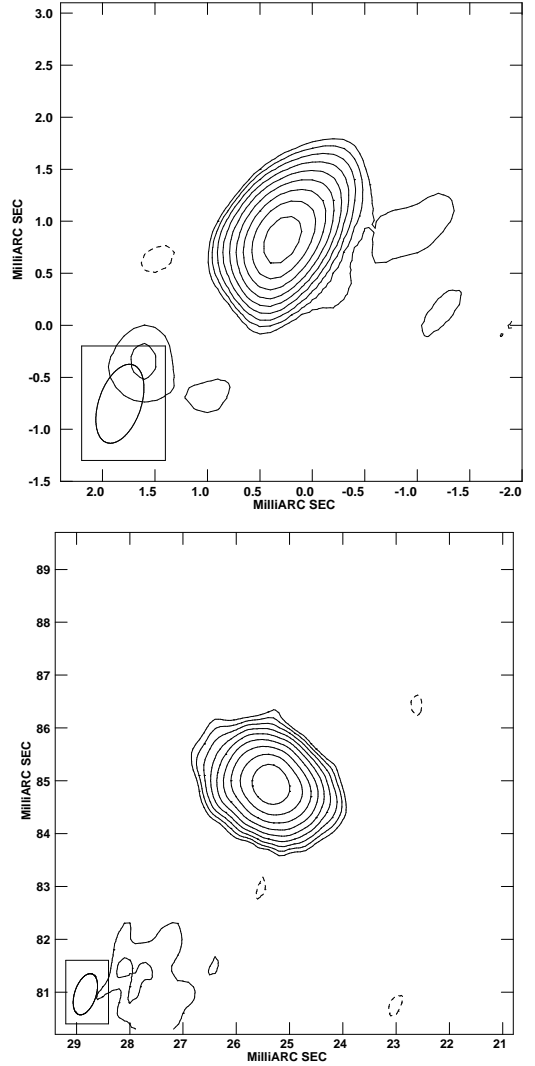
 
\centering 
\includegraphics[width=7cm, angle=-90]{j2007.eps} 
\includegraphics[width=6.8cm]{2033.eps} 
\caption{Phase-referenced images of the background sources J2007
({\it top}) and J2033 ({\it bottom}) from the second epoch of the VLBA observations.
The images are in milliarcseconds offset to the positions in Table
\ref{ta:sources}. The
contours start at a 3$\sigma$ level, namely $5.1\times10^{-2}$ and
$1.2\times10^{-3}\,\mathrm{Jy\,beam^{-1}}$, respectively, and increase
by $\sqrt{2}$. The first negative contour ($-3\sigma$) is shown by dashed contours. The synthesized beam is shown in the bottom left corner. \label{fig:qsoima_afgl}}                                 
\end{figure}

\begin{figure} 
\center 
\includegraphics[width=7cm, angle=-90]{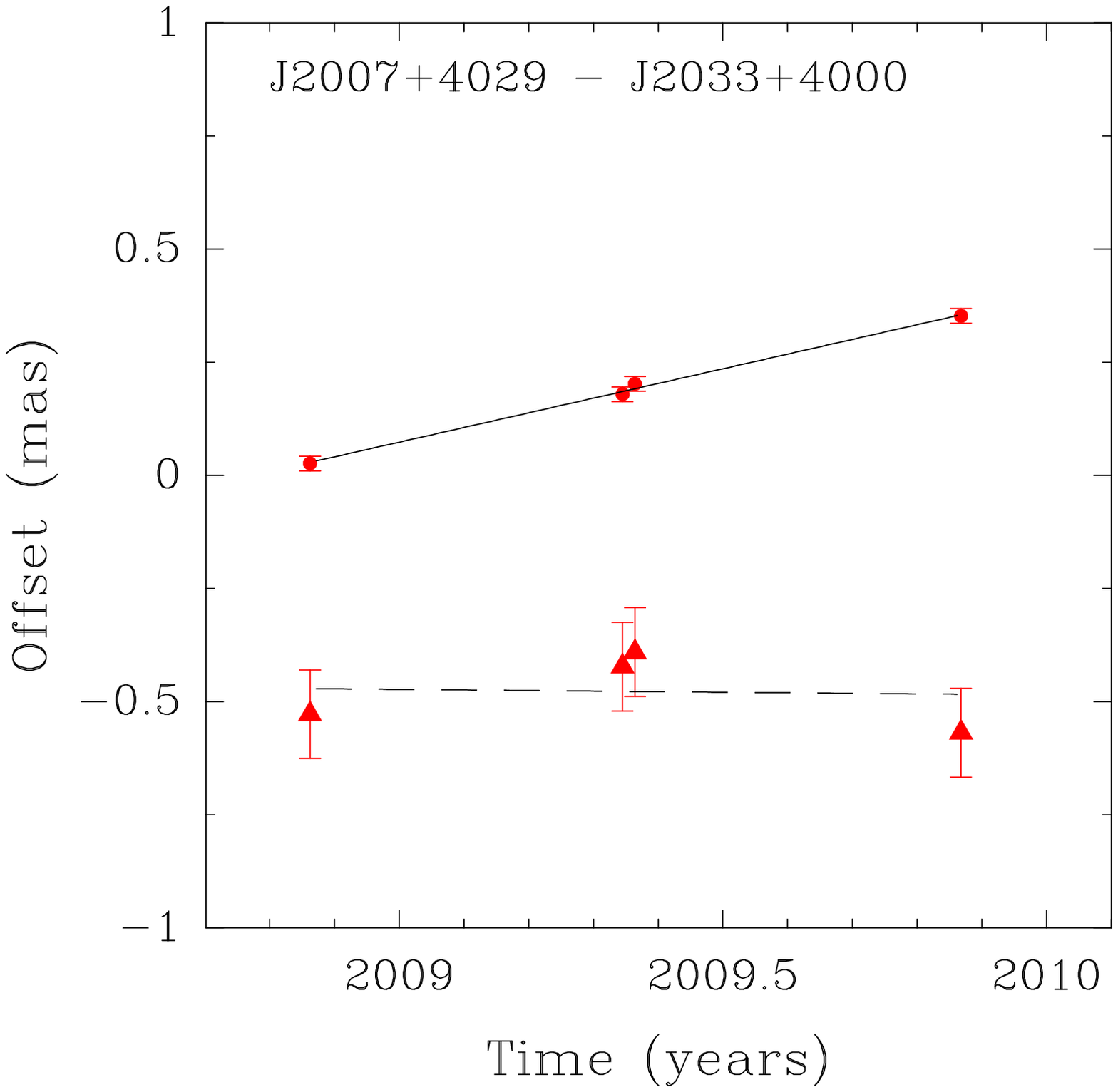} 
\caption{\label{fig:qso_afgl} Variation in time of the relative positions
of background source J2033 with respect to J2007.
The dots mark the right ascension data points, while the
filled triangles represent the declination data points. The solid
line shows the right ascension fit, while the dashed line represents
the declination fit.}                                                   
\end{figure} 

\label{parfit}

The VLBA 22 GHz observations included four background sources, of which
two were used for astrometry, J2007 and J2033 (Fig.\,\ref{fig:qsoima_afgl}).                                           
Two background sources were discarded, one because of a non-detection
(J2033+4040) and the other one because of a strong structure change during the observations (J2032+4057). 
A third source (J2033) was found to be heavily resolved by the VLBA
(see Fig.\,\ref{fig:qsoima_afgl}). For this quasar, the position was derived by fitting only the peak of the emission in
the image, instead of its whole spatial structure.
 
The parallax and proper motion fitting of the water maser was done in the same fashion as for the methanol maser data.                    
As the VLBA observations used two background sources, the ``averaged
data'' were calculated separately for each background source. 
The combined fit was carried
out on the two averaged data sets, and the parallax uncertainty was
multiplied by $\sqrt N$ for the number of maser spots to account for correlated differential positions (see above). In the combined
fit, we treated the proper motion of the maser separately from the two background
sources, taking their relative, apparent (linear) motion into account (Fig.\,\ref{fig:qso_afgl}). 
This is necessary, because there was a significant apparent linear motion in the right
ascension coordinate ($0.32\pm0.01\,\mathrm{mas\,yr^{-1}}$), but
none in declination ($-0.01\pm0.1\,\mathrm{mas\,yr^{-1}}$).
 
With the parallax and averaged proper motion results we calculated
the 3D space velocities of the SFRs with respect to the Galactic center (\citealt{reid:2009b}). Each SFR has three velocity vectors: $U$, the velocity in the direction of the Galactic
center; $V$, the velocity in the direction of the Galactic rotation; and $W$, the velocity in
the direction of the North Galactic Pole (NGP). The maser proper motions that we measure are the velocity differences between the maser source and the Sun. For obtaining the 3D space velocity, we first need to subtract the solar motion and then to transform the velocity vector from a nonrotating frame to a frame rotating with Galactic rotation. 
In this calculation we used the solar  peculiar motion obtained by \cite{schoenrich:2010} and
assumed a flat Galactic rotation curve with $\theta=239\,\mathrm{km~s^{-1}}$
and a solar distance to the Galactic center $R=8.3$\,kpc. These are the revised Galactic rotation parameters derived by \citet{brunthaler:2011} taking the recent revision
of the solar peculiar motion into account, from the {\it Hipparcos} values ($U_\odot$, $V_\odot$, $W_\odot$) = (10.00, 5.25, 7.17) $\mathrm{km~s^{-1}}$ by \cite{dehnen:1998}  to ($U_\odot$, $V_\odot$, $W_\odot$) = (11.10, 12.24, 7.25) $\mathrm{km~s^{-1}}$ by \cite{schoenrich:2010}.

\section{Results} 
 
\begin{table*} 
\centering 
\caption{ Parallax, proper motions, and space velocities toward Cygnus~X
\label{ta:parallax}}                                                   
\begin{tabular}{l l l l l r r r r} 
\hline\hline 
\noalign{\smallskip} 
Source & \multicolumn{1}{c}{Parallax} &  \multicolumn{1}{c}{$\mathrm{D_{Sun}}$} &  \multicolumn{1}{c}{$<\mu_\alpha>$} &  \multicolumn{1}{c}{$<\mu_\delta>$}& \multicolumn{1}{c}{$v_{\mathrm{LSR}}$}& \multicolumn{1}{c}{$U$}& \multicolumn{1}{c}{$V$}& \multicolumn{1}{c}{$W$} \\                                       
              &  \multicolumn{1}{c}{(mas)} &  \multicolumn{1}{c}{(kpc)} &  \multicolumn{1}{c}{(mas~yr$^{-1}$)} &   \multicolumn{1}{c}{(mas~yr$^{-1}$)}& \multicolumn{1}{c}{(km~s$^{-1}$)} &  \multicolumn{1}{c}{(km~s$^{-1}$)} &  \multicolumn{1}{c}{(km~s$^{-1}$)} &  \multicolumn{1}{c}{(km~s$^{-1}$)}\\        
\hline\hline 
\noalign{\smallskip} 
W\,75N         &$0.772\pm0.042$ & $1.30^{+0.07}_{-0.07}$ & $-1.97\pm0.10$&$-4.16\pm0.15$&$9.0^1$&$-0.5\pm1.0$&$3.6\pm2.5$&$1.2\pm0.8$\\
\noalign{\smallskip} 
DR\,21          & $0.666\pm0.035$ & $1.50^{+0.08}_{-0.07}$ &$-2.84\pm0.15$&$-3.80\pm0.22$&$-3.0^1$&$-0.0\pm1.5$&$-8.3\pm2.5$&$6.6\pm1.3$ \\   
\noalign{\smallskip} 
DR\,20          &$0.687\pm0.038$ & $1.46^{+0.09}_{-0.08}$&$-3.29\pm0.13$&$-4.83\pm0.26$&$-3.0^2$&$7.5\pm1.5$&$-8.8\pm2.5$&$5.1\pm1.3$
\\                                                                      
\noalign{\smallskip} 
IRAS\,20290+4052 & $0.737\pm0.062$&$1.36^{+0.12}_{-0.11}$&$-2.84\pm0.09$&$-4.14\pm0.54$&$-1.4^3$&$1.9\pm1.8$&$-8.2\pm2.5$&$6.0\pm2.1$\\
\noalign{\smallskip} 
AFGL\,2591 & $0.300\pm0.010$ & $3.33^{+0.11}_{-0.11}$& $-1.20\pm0.32$
& $-4.80\pm0.12$ &$-5.7^4$&$-12.8\pm3.4$&$-10.4\pm2.5$&$-22.1\pm4.3$\\
\noalign{\smallskip} 
\hline 
\end{tabular} 
\tablebib{(1) \citet{dickel:1978,schneider:2010}, (2) \citet{motte:2007}, (3) \citet{bronfman:1996}, (4) \citet{tak:1999}.}
\end{table*} 
 
Table~\ref{ta:parallax} shows the parallax and averaged proper motion
results for all SFRs and their calculated space velocities, while
detailed results of the parallax and proper motion fitting are found
in Table~\ref{ta:parallax_details}.                                     
 
\begin{figure*} [!htbp]
\center 
\includegraphics[width=14cm]{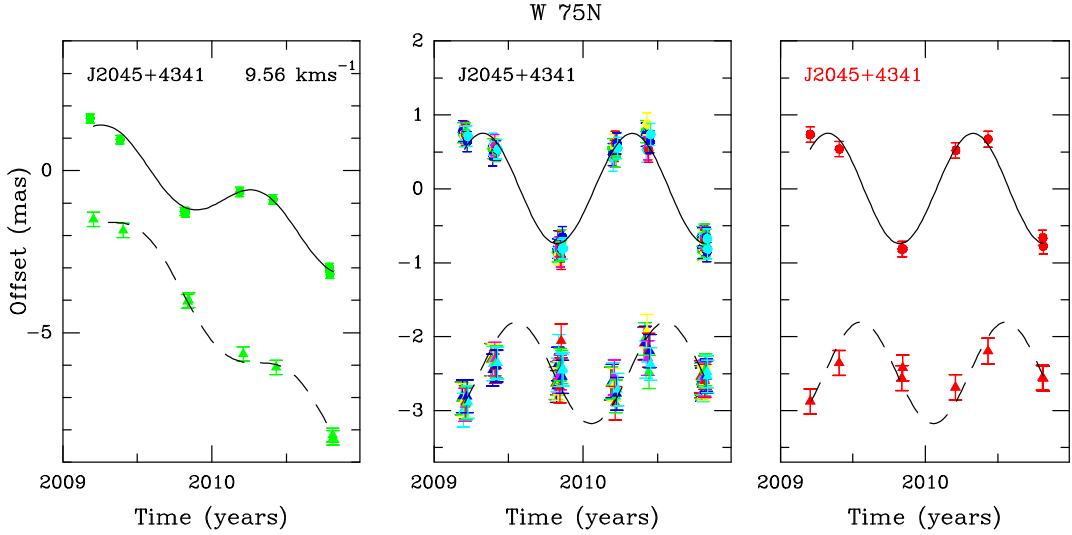}
\caption{Results of the parallax fit for W\,75N. {\it Left panel:} Results of the parallax fit for the maser spot at $V_\mathrm{LSR}=9.56\,\mathrm{km\,s^{-1}}$ without the removal of the proper motions. The right ascension and declination data has been offset for clarity {\it Middle and right panel:} Results of the parallax fit for W\,75N after removing proper motions and positional offsets. The declination data has been offset for clarity.
Combined fit on 10 maser spots with respect to J2045 ({\it middle panel}).  Fit to the averaged data of the 10 maser spots ({\it right panel}). The dots mark the data points in right ascension, while the filled triangles
mark the declination data points. The solid lines show the resulting
parallax fit in right ascension and the dashed lines show the fit
in declination. The scale of the Y-axis of the {\it middle} and {\it right panels} is different from the scale of the {\it left panel}. \label{fig:par_w75}}                                    
\end{figure*} 
 
\subsection{Cygnus~X North: W\,75N, DR\,21, and DR\,20} 
 
W\,75N has 14 different methanol maser features emitting in an local standard of rest velocity ($V_\mathrm{LSR}$)
range of  3--10 $\mathrm{km~s^{-1}}$ with 32 maser spots in total 
(see Fig.\,\ref{fig:masers}). After removing the maser spots with
nonlinear motions, we were left with ten spots belonging to six maser
features for the parallax and proper motion fitting. Figure~\ref{fig:par_w75}
shows the parallax fit of W\,75N, resulting in $0.772\pm0.042$\,mas
or $1.30^{+0.07}_{-0.07}$ kpc. While we usually plot the parallax fits after removing the proper motion, in this figure we also show a parallax fit, to one of the maser spots of W\,75N, without the removal of the proper motion. Since different maser spots can have different proper motions, it is not instructive to plot the parallax fits to all the maser spots when including the proper motion.

\begin{figure*} 
\center 
\includegraphics[width=14cm]{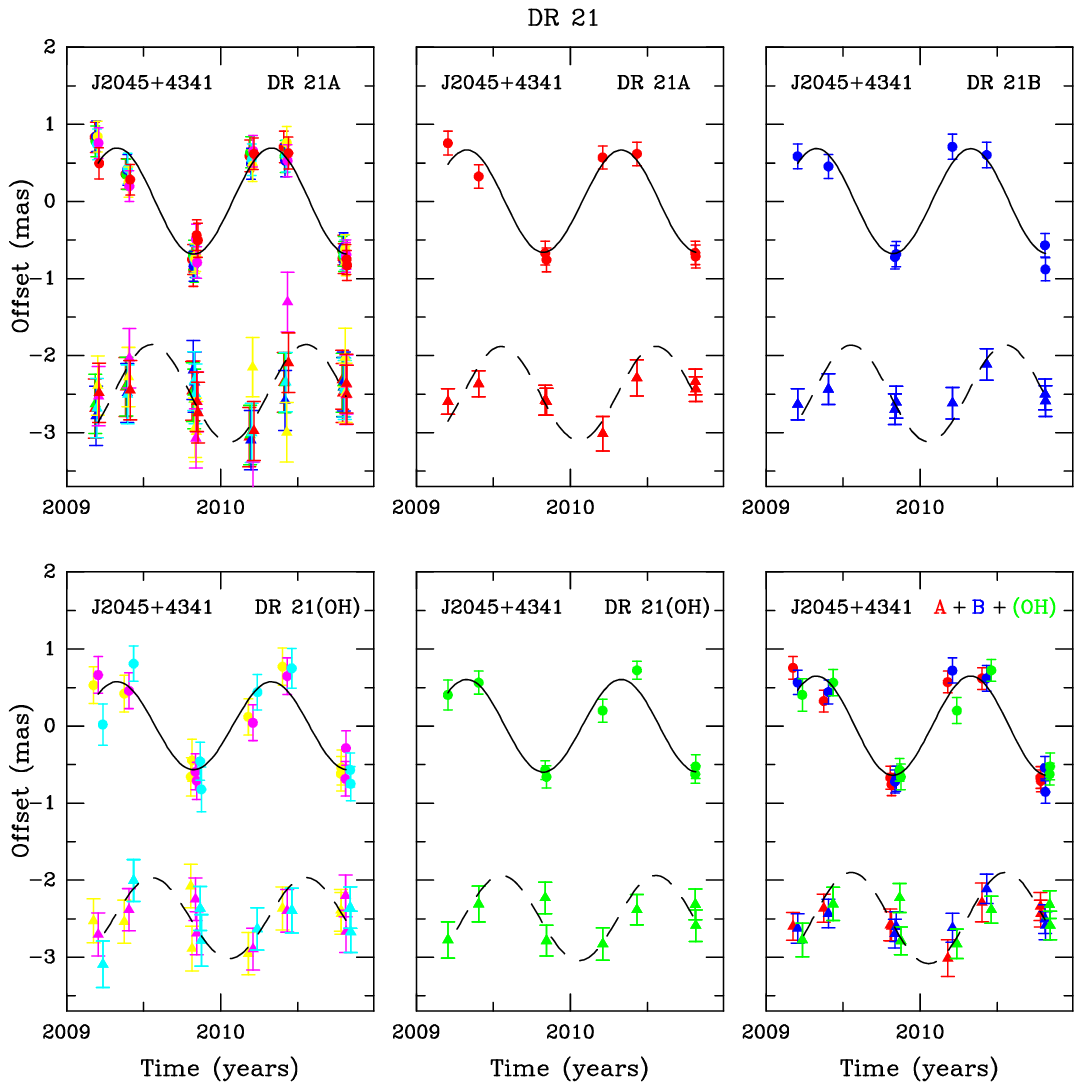} 
\caption{Results of the parallax fit for DR\,21 after removing proper motions and positional offsets. The declination data has been offset for clarity. {\it Top row, from left to
right, first panel:} Combined fit on 7 maser spots from DR\,21A with
respect to J2045. {\it Second panel:} Fit to the averaged data
of the 7 maser spots of DR\,21A. {\it Third panel:} Single fit to
DR\,21B maser spot at  3.96\,$\mathrm{km~s^{-1}}$. {\it Bottom row, from left to right, first panel:}
Combined fit on 3 maser spots in DR\,21(OH). {\it Second panel:} Fit to the averaged data of the 3 maser spots of DR\,21(OH). {\it Third panel:}
Combined fit on the averaged data set of DR\,21A, DR\,21(OH), and the single maser
spot of DR\,21B. The dots mark the data points in right ascension,
while the filled triangles mark the declination data points. The
solid lines show the resulting parallax fit in right ascension and
the dashed lines show the fit in declination. \label{fig:par_dr21}}     
\end{figure*} 
 
For DR\,21A we found 18 maser spots belonging to seven maser features,
covering an LSR velocity range of $-9.4$ to $-5.2\,\mathrm{km~s^{-1}}$
(see Fig.\,\ref{fig:masers}). The parallax fitting was based on seven
spots in four maser features. Toward DR\,21B we found seven spots in three
maser features with V$_\mathrm{LSR}$ between 3.3 and 5.0\,$\mathrm{km~s^{-1}}$
(see Fig.\,\ref{fig:masers}), but only one maser spot had a good
parallax fit.  We also detected methanol masers in DR\,21(OH) where we used three maser spots with V$_\mathrm{LSR}$=$-2.5$ to $-3.5\,\mathrm{km~s^{-1}}$ for fitting a parallax signature. 
If we assume that the masers in DR\,21A, DR\,21B, and DR\,21(OH) belong to the same SFR -- which is likely since their parallaxes are consistent: $0.686\pm0.060$ for DR\,21A, $0.705\pm0.072$ for DR\,21B, and $0.622\pm0.055$ for DR\,21(OH) -- the resulting parallax fit for DR\,21 becomes $0.666\pm0.035$\,mas, or $1.50^{+0.08}_{-0.07}$ kpc. The connection between DR\,21 and DR\,21(OH) has already been noted by \citet{schneider:2006}, since these SFRs are part of the same elongated filament seen in CO emission.
The parallax fits to the combined and averaged data for DR\,21A and DR\,21(OH), the fit to the individual spot of DR\,21B, and their overall combination (i.e., for the DR\,21 SFR) are shown in Fig.\,\ref{fig:par_dr21}.

\begin{figure}
\includegraphics[width=9cm]{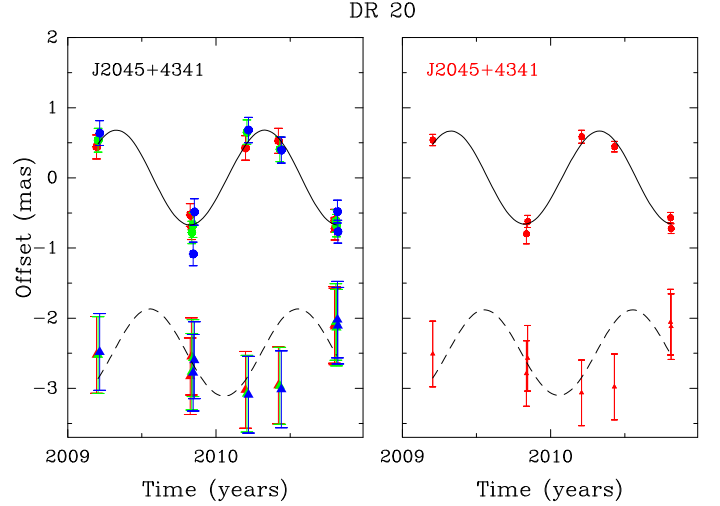} 
\caption{Results of the parallax fit for DR\,20 after removing proper motions and positional offsets. The declination data has been offset for clarity. {\it Left panel:}
Combined fit on 3 maser spots with respect to J2045. {\it Right
panel:} Fit to the averaged data of the 3 maser spots. The dots mark the data points in right ascension, while the filled triangles
mark the declination data points. The solid lines show the resulting
parallax fit in right ascension and the dashed lines show the fit
in declination. \label{fig:par_dr20}}                                   
\end{figure} 
 
Six maser spots in the velocity range of $-4.8$ to $-2.0\,\mathrm{km~s^{-1}}$,
belonging to two maser features that were found in DR\,20 (see Fig.\,\ref{fig:masers}).
The parallax fit was based on one maser feature with three maser
spots. We find a parallax of $0.687\pm0.038$\,mas, corresponding to
a distance of $1.46^{+0.10}_{-0.09}$ kpc (see the parallax fit in Fig.\,\ref{fig:par_dr20}).
The second epoch was omitted from the parallax fitting, because the maser spots in DR\,20 had an outlying data point in proper motion caused by large residuals in atmospheric delay in that  epoch's data.

\subsection{Cygnus~X South: AFGL\,2591 and IRAS\,20290+4052} 
 
\begin{figure}
\centering 
\includegraphics[width=9cm]{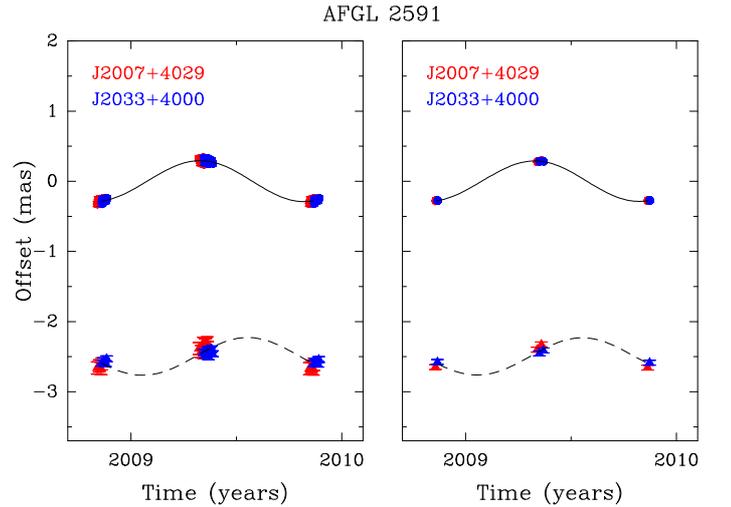} 
\caption{Results of the 22 GHz water maser parallax fit for AFGL\,2591 (VLBA data) after removing the proper motions and positional offsets. The declination data has been offset for clarity. 
{\it Left panel:} Combined fit on 6 maser spots with respect to J2007 and J2033. {\it Right
panel:} Fit to the averaged data of the 6 maser spots. The dots mark the data points in right ascension, while the filled triangles
mark the declination data points. The solid lines show the resulting
parallax fit in right ascension and the dashed lines show the fit
in declination. \label{fig:p_afgl}}                                                      
\end{figure}

The parallax of AFGL\,2591 was based on VLBA data using 22 GHz water masers. 
AFGL\,2591 has a very rich water-maser spectrum of 80 maser features
with V$_\mathrm{LSR}$ from $-34$ to $-0.4\,\mathrm{km~s^{-1}}$ (see \citealt{sanna:2011}). The parallax and proper motion fits were
based on six maser features and resulted in a parallax of $0.300\pm0.010$\,mas
or $3.33^{+0.11}_{-0.11}$ kpc. Figure~\ref{fig:p_afgl} shows the parallax and proper
motion fit of AFGL\,2591 with respect to the two background sources. The parallax fit to each background source separately was $0.302\pm0.009$\,mas (J2007) and $0.299\pm0.002$\,mas (J2033).
The difference between the two parallax fits is due to the apparent motion observed between the two background sources. The effect on the parallax can be approximated by the difference between the two results, namely 0.003\,mas.
The previous distance measurements of AFGL\,2591 were near 1.6\,kpc. Assuming that the SFR is associated to IC\,1318c (\citealt{wendker:1974}), of which the distance was determined by \cite{dickel:1969}, AFGL\,2591 was thought to be at 1.5\,kpc. Alternatively, \cite{dame:1985} suggest a distance of 1.7\,kpc based kinematic distances to CO clouds. However, one can find a much wider spread of distances from 1 to 2\,kpc in the literature (\citealt{tak:1999}). 
The water maser parallax puts AFGL\,2591 a factor 2.22 (assuming 1.5\,kpc) farther away, at 3.33\,kpc, which implies a dramatic change for the physical properties of the SFR (discussed in \citealt{sanna:2011}).

\begin{figure} 
\center 
\includegraphics[width=4.5cm]{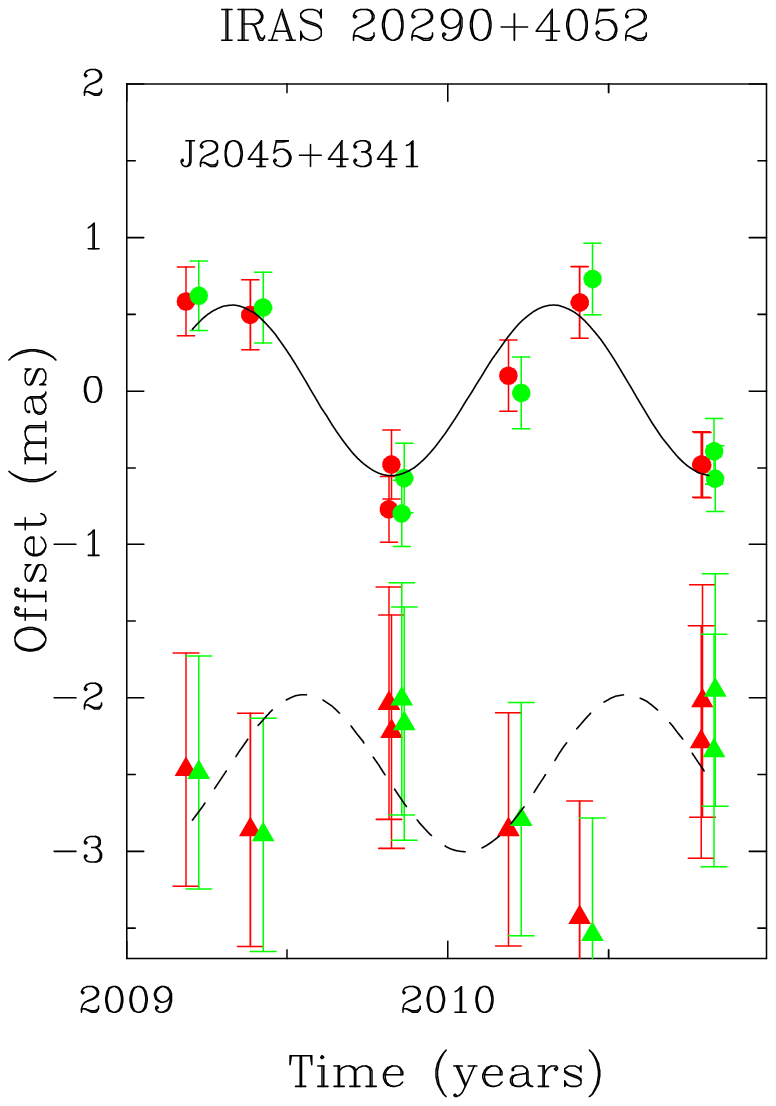} 
\caption{Results of the parallax fit for IRAS\,20290 after removing proper motions and positional offsets. The declination data has been offset for clarity. Shown is
the combined fit on 2 maser spots with respect to J2045. The dots mark the data points in right ascension, while the filled
triangles mark the declination data points. The solid lines show
the resulting parallax fit in right ascension and the dashed lines
show the fit in declination. \label{fig:par_ob2}}                       
\end{figure}

Finally, for IRAS\,20290, we found two methanol maser features composed
of a total of eight spots with V$_\mathrm{LSR}$ from $-6.2$ to $-3.8\,\mathrm{km~s^{-1}}$
(see Fig.\,\ref{fig:masers}). Due to large residuals for the fit,
only two maser spots, belonging to the same feature, were suitable
for parallax fitting resulting in a parallax of $0.737\pm0.062$\,mas
or $1.36^{+0.12}_{-0.11}$\,kpc. The parallax fit is shown in Fig.\,\ref{fig:par_ob2}. 

\onecolumn 
\begin{table} 
\centering 
\caption{Detailed results of parallax and proper motions measurements
\label{ta:parallax_details}}                                            
\begin{tabular}{l c c  l l l} 
\hline 
\hline 
Maser & Background &  $v_\mathrm{LSR}$ &  \multicolumn{1}{c}{Parallax} & \multicolumn{1}{c}{$\mu_\alpha$} & \multicolumn{1}{c}{$\mu_\delta$}\\                                                        
 & source     & ($\mathrm{km~s^{-1}}$)    & \multicolumn{1}{c}{(mas)}    & \multicolumn{1}{c}{($\mathrm{mas~yr^{-1}}$)}& \multicolumn{1}{c}{($\mathrm{mas~yr^{-1}}$)}\\                                            
\noalign{\smallskip} 
\hline 
\noalign{\smallskip} 
{\bf W\,75N} & J2045+4341	& 9.56      &$0.750\pm0.070~$&$-2.01\pm0.13$ &$-4.33\pm0.10$\\
&			& 9.23     &$0.800\pm0.067~$&$-1.99\pm0.11$ &$-4.37\pm0.11$\\
&			& 9.23     &$0.730\pm0.079~$&$-2.03\pm0.13$ &$-4.37\pm0.15$\\
&			& 9.23     &$0.748\pm0.062~$&$-1.85\pm0.09$ &$-4.09\pm0.19$\\
 &  			& 8.88     &$0.879\pm0.060~$&$-1.97\pm0.09$ &$-4.42\pm0.13$\\
 &                           & 8.88     &$0.733\pm0.068~$&$-1.87\pm0.11$&$-4.11\pm0.13$\\
& 			& 6.77     &$0.764\pm0.068~$&$-2.18\pm0.10$ &$-4.17\pm0.20$\\
&			& 4.31     &$0.725\pm0.072~$&$-1.99\pm0.11$ &$-3.93\pm0.17$\\
&			& 3.26      &$0.746\pm0.056~$&$-1.89\pm0.04$ &$-3.94\pm0.12$\\
& 			& 2.90      &$0.801\pm0.056~$&$-1.96\pm0.08$ &$-3.85\pm0.17$\\
&&\multicolumn{1}{l}{Combined fit}&$0.772\pm0.063$ &\multicolumn{1}{c}{--}&\multicolumn{1}{c}{--}\\ 
&			 &\multicolumn{1}{l}{Averaging data}&$\mathbf{0.772\pm0.042}~$&\multicolumn{1}{c}{--}&\multicolumn{1}{c}{--}\\                                                                
&$<\mu> $			  &&&$-1.97\pm 0.10$&$-4.16\pm 0.15$\\                                                                
\hline
\noalign{\smallskip} 
{\bf DR\,21A}&J2045+4341& $-5.54$   &$0.751\pm 0.094~$&$-2.67\pm0.15$ &$-3.22\pm0.18$\\
&& $-5.89$   &$0.706\pm0.087~$&$-2.76\pm0.14$ &$-3.12\pm0.15$\\
&& $-6.59$   &$0.709\pm0.094~$&$-2.94\pm0.15$ &$-2.94\pm0.24$\\
&& $-6.94$   &$0.718\pm0.068~$&$-2.87\pm0.11$ &$-2.98\pm0.17$\\
&& $-9.05$   &$0.710\pm0.11~$&$-2.59\pm0.17$ &$-3.87\pm0.38$\\
&& $-9.05$   &$0.719\pm0.11~$&$-2.78\pm0.17$ &$-3.67\pm0.35$\\
&& $-9.40$   &$0.600\pm0.088~$&$-3.26\pm0.14$ &$-4.37\pm0.14$\\
&			 &\multicolumn{1}{l}{Combined fit}&$0.713\pm0.091$ &\multicolumn{1}{c}{--}&\multicolumn{1}{c}{--}\\ 
&			 &\multicolumn{1}{l}{Averaging data}& $\mathbf{0.686\pm0.060}~$&\multicolumn{1}{c}{--}&\multicolumn{1}{c}{--}\\                                                                
&$<\mu> $& &&$-2.84\pm0.15$&$-3.56\pm 0.25$\\
\noalign{\smallskip} 
{\bf DR\,21B} &J2045+4341  	&3.96    &$\mathbf{0.705\pm0.072}~$&$-2.88\pm0.11$&$-4.30\pm0.13$\\
\noalign{\smallskip} 
{\bf DR\,21(OH)} &J2045+4341    & $-2.72$ & $0.694\pm0.100$ & $-2.31\pm0.16$ &$-4.29\pm0.16$\\
&                            &$-3.07$ & $0.533\pm0.111$ & $-3.11\pm0.19$ & $-4.51\pm0.17$\\
&                            &$-3.43$ & $0.572\pm0.104$ & $-3.05\pm0.16$ & $-4.51\pm0.22$\\
&			 &\multicolumn{1}{l}{Combined fit}&$0.591\pm0.104$ &\multicolumn{1}{c}{--}&\multicolumn{1}{c}{--}\\ 
&			 &\multicolumn{1}{l}{Averaging data}& $\mathbf{0.622\pm0.055}~$&\multicolumn{1}{c}{--}&\multicolumn{1}{c}{--}\\           
{\bf DR\,21}\tablefootmark{a}&J2045+4341   &  &$\mathbf{0.666\pm0.035}$ &\multicolumn{1}{c}{--}&\multicolumn{1}{c}{--}\\ 
&$<\mu> $				 & &&$-2.84\pm0.15$&$-3.80\pm 0.22$\\
\hline
\noalign{\smallskip} 
{\bf DR\,20}&J2045+4341 	& $-3.78$     &$0.634\pm0.044~$&$-3.16\pm0.06$ &$-4.82\pm0.39$\\
	&		& $-4.13$    &$0.729\pm0.056~$&$-3.41\pm0.08$ &$-4.85\pm0.42$\\
&			& $-4.48$   &$0.726\pm0.136~$&$-3.29\pm0.21$ &$-4.81\pm0.43$\\
&			&\multicolumn{1}{l}{Combined fit}&$0.700\pm0.085$ &\multicolumn{1}{c}{--}&\multicolumn{1}{c}{--}\\ 
&			 &\multicolumn{1}{l}{Averaging data}& $\mathbf{0.687\pm0.038}~$&\multicolumn{1}{c}{--}&\multicolumn{1}{c}{--}\\                                                                
&$<\mu> $				 & &&$-3.29\pm0.13$&$-4.83\pm0.26$\\	                                                              
\hline
\noalign{\smallskip} 
{\bf IRAS\,20290+4052 } & J2045+4341      & $-5.54$    &$0.691\pm0.060~$&$-2.82\pm0.09$&$-4.17\pm0.50$\\
&		        & $-5.89$    &$0.785\pm0.063~$&$-2.85\pm0.09$ &$-4.11\pm0.57$\\
&                          &\multicolumn{1}{l}{Combined fit}&$\mathbf{0.737\pm0.062}$&\multicolumn{1}{c}{--}&\multicolumn{1}{c}{--}\\                                                                
&$<\mu> $				 &&&$-2.84\pm0.09$&$-4.14\pm0.54$ \\                                                              
\hline
\noalign{\smallskip} 
{\bf AFGL\,2591} &J2007+4029 &--4.58        & $ 0.267 \pm 0.009$    & $ -0.656\pm 0.025$ & $ -5.906 \pm 0.051$  \\                                    
&  &--5.00  & $ 0.308 \pm 0.014$    & $ -1.083 \pm 0.037$ & $-4.019 \pm 0.169$  \\                                                   
 & & --5.84  & $ 0.290 \pm 0.007$    & $ -1.271 \pm 0.018$ & $-4.152 \pm 0.169$  \\                                                   
 &&--7.53  & $ 0.317 \pm 0.004$    & $ -0.955 \pm 0.011$ & $ -4.715\pm0.071$  \\                                                              
  &&--7.95  & $ 0.280 \pm 0.005$    & $ -1.385 \pm 0.015$ & $-5.745 \pm 0.155$  \\                                                   
 & &--10.90  &$ 0.348 \pm 0.008$    & $ -0.884 \pm 0.021$ & $-4.398 \pm 0.113$  \\                                                   
  &                        &\multicolumn{1}{l}{Averaging data}& $0.302 \pm 0.009$    & \multicolumn{1}{c}{--}&\multicolumn{1}{c}{--}\\
&J2033+4000 &--4.58   & $ 0.263 \pm 0.003$    & $ -0.980 \pm 0.009$& $ -5.863 \pm 0.169$  \\                                               
 & &   --5.00      & $ 0.301 \pm 0.009$    & $ -1.407 \pm 0.025$& $ -3.976 \pm 0.025$  \\                                               
 & &  --5.84  & $ 0.287 \pm 0.001$    & $ -1.595 \pm 0.002$ &$ -4.109 \pm 0.028$  \\                                                 
&    &  -7.53& $ 0.315 \pm 0.001$    & $ -1.279 \pm 0.004$ & $-4.672 \pm 0.141$  \\                                                   
& &  --7.95& $ 0.277 \pm 0.001$    & $ -1.709 \pm 0.001$ & $ -5.702\pm 0.031$  \\                                                          
& & -10.90 & $ 0.345 \pm 0.002$    & $ -1.208 \pm 0.005$ & $ -4.355\pm 0.078$  \\                                                          
  &                        &\multicolumn{1}{l}{Averaging data}&$0.299 \pm 0.002$    & \multicolumn{1}{c}{--}&\multicolumn{1}{c}{--}\\
&Both QSOs              &\multicolumn{1}{l}{Combined fit}& $ \mathbf{0.300\pm 0.010}$    & \multicolumn{1}{c}{--}&\multicolumn{1}{c}{--}\\
 &$<\mu> $				 & &&$-1.20\pm0.32$&$-4.80\pm0.12$ \\                                                              
\noalign{\smallskip} 
\hline 
\hline 
\end{tabular} 
\tablefoot{
\tablefoottext{a}The resulting parallax fit for DR\,21 is based on the averaged data of DR\,21A, DR\,21(OH), and the single maser spot of DR\,21B. 
}
\end{table} 
\twocolumn

\section{Is Cygnus~X one region?} 
 
\begin{figure} 
\centering 
\includegraphics[width=9cm]{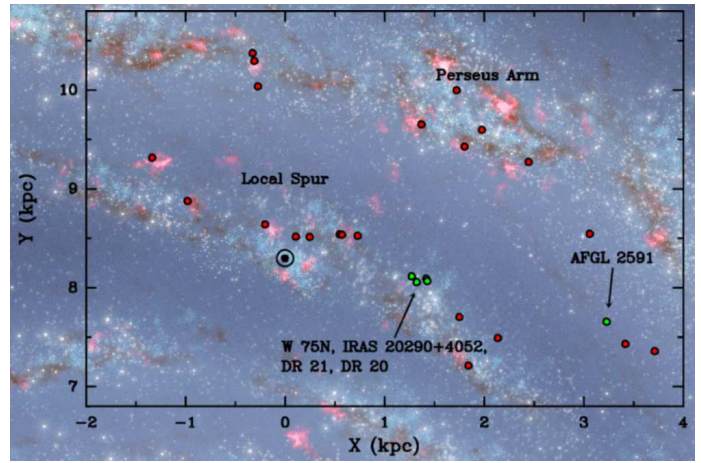}
\caption{\label{fig:LA} An artist's impression of the Galactic plane (image credit: R.~Hurt, NASA/JPL-Caltech/SSC) as seen from the North Galactic Pole with the Galactic center (not shown) located at (0,0) kpc. The Sun is located at (0, 8.3) kpc and marked by a black circle around a black dot.
Dots mark the sources to which an accurate distance
is known in the Local Arm: green dots mark sources from this work; red dots mark sources from the literature (from left to right): \citet{choi:2008}, \citet{reid:2009a}, \citet{hirota:2007,menten:2007}, \citet{hirota:2008a}, \citet{hirota:2011}, \citet{moscadelli:2009}, \citet{rygl:2010a}, \citet{hirota:2008b}, \citet{reid:2011},  \citet{xu:2009a}, \citet{rygl:2010a,nagayama:2011}, Xu, private comm., and \citet{ando:2011}. For a comparison we also plot the sources in the Perseus Arm (from left to right): \citet{reid:2009a}, \citet{niinuma:2011}, \citet{oh:2010} , \citet{xu:2006, hachisuka:2006}, \citet{asaki:2010}, \citet{moellen:2007}, \citet{sato:2008,rygl:2010a}, \citet{moscadelli:2009}, and \citet{oh:2010}. The Local and Perseus Arms
are indicated.}                                                            
\end{figure}

\begin{figure*} 
\centering
\includegraphics[angle=-90,width=18cm]{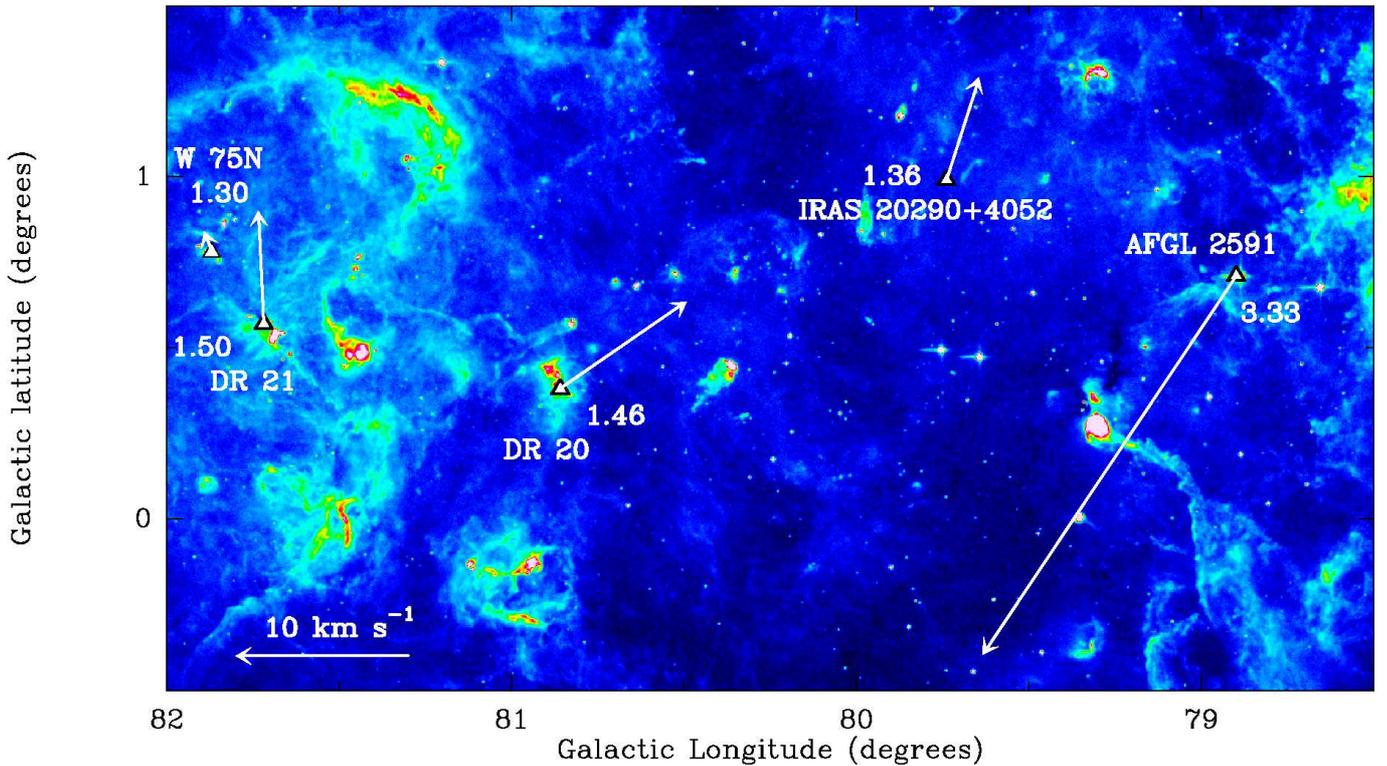} 
\caption{\label{fig:small}  {\it Midcourse Space eXperiment} (MSX) 8\,$\mu$m image of the Cygnus~X region overlaid with the resulting $UVW$ space motions for each source. The white triangles mark the water maser (AFGL\,2591) and the methanol maser sources, with their distances indicated in kpc. }
\end{figure*}

AFGL\,2591, supposedly located in the Cygnus~X South region, has
a much farther distance than previously assumed, which implies
that AFGL\,2591 is not a part of a single Cygnus~X complex. 
AFGL\,2591 is perhaps part of the Local Arm, which would then extend to greater distances from
the Sun than currently thought. While we need more data to investigate this issue with
stronger statistical support, we note that AFGL 2591 is not the only source found in this region
(see Fig.\,\ref{fig:LA}). Two recent parallax measurements, namely ON2 at 3.83 kpc (\citealt{ando:2011})
and G75.76+0.35 at 3.37 kpc (Xu, private comm.; see also the Bar and Spiral Structure Legacy
survey, BeSSeL, website\footnote{http://www.mpifr-bonn.mpg.de/staff/abrunthaler/BeSSeL/index.shtml}), locate these SFRs close in space to AFGL 2591 (within 3$^\circ$
on the sky).
AFGL\,2591 has a $V_\mathrm{LSR}$ similar to other SFRs in Cygnus~X North, even though it is projected against the Cygnus complex and not part of it.
The space motions of AFGL\,2591, though, are very different from the sources
in Cygnus~X as can be seen in Fig.\,\ref{fig:small}. 
We note that the distance measurement of AFGL\,2591 is not affected by using different VLBI arrays and maser transitions than for the other SFRs. Maser parallaxes, both the water and the 6.7 (and 12.2) GHz methanol masers have been shown to produce robust distance measurements: for example, the distance to W3(OH), measured with both 12.2 GHz methanol (\citealt{xu:2006}) and 22 GHz water ({\citealt{hachisuka:2006}}) masers, yielded the same number, whereas the 6.7 GHz maser distances (\citealt{rygl:2010a}) also agreed with the VERA water maser distances (\citealt{sato:2008, nagayama:2011}).

The most important result of this study is that Cygnus~X North is one physically related complex of SFRs, including W\,75N, DR\,21, DR\,20, and IRAS\,20290 (and therefore probably also the Cygnus OB\,2 association), located at $1.40^{+0.08}_{-0.08}$\,kpc. This is an average of the individual distances, which range from 1.30 to 1.50\,kpc. Our data are consistent with a single distance for these sources, within measurement uncertainty. We note that our distance to the Cygnus~X complex is similar to the photometric distance of 1.5\,kpc obtained by \cite{hanson:2003} toward the Cyg OB\,2 association.
Compared to the extent on the sky of the SFRs mentioned above, $25\times60$\,pc, the distance spread of 200\,pc is a factor $\sim$3.5 wider (however, the measurement uncertainty is a factor 1--2 times the angular extent).
The parallax results can possibly be extended based on contingent spatial-velocity structures identified by \citet{schneider:2006} in the CO emission. Following these authors DR\,22, DR\,23, DR\,17, and AFGL\,2620 (their CO groups I and II) should also be part of Cygnus~X North and thus at the same distance.

Our results did not find any evidence of a southern counterpart to Cygnus~X North, since AFGL\,2591 was found to be much more distant and other SFRs toward Cygnus~X South (except for IRAS\,20290 at $l$=79$\rlap{$.$}\,^\circ7$, which was found to be at the same distance as Cygnus~X North) were not included in this work. \citet{schneider:2006} find that most of the mass in Cygnus~X South is contained in a group of molecular clouds: DR\,4, DR\,5, DR\,12, DR\,13, and DR\,15 (their CO group IV). Parallax measurements to one of these SFRs then could confirm that Cygnus~X South is connected to Cygnus~X North, thus forming one of the largest known giant molecular cloud structures in the Milky Way of $\sim2.7\times10^6\,\mathrm{M_\odot}$ (\citealt{schneider:2006} rescaled to 1.4\,kpc).

The distances of the SFRs in Cygnus~X fit well with the trajectory
of the Local Arm between 0.5 and 2.5 kpc in the X coordinate (see
Fig.\,\ref{fig:LA}), defined by measurements to Cep\,A (\citealt{moscadelli:2009}),
L\,1206 (\citealt{rygl:2010a}), L\,1448C (\citealt{hirota:2011}), Cygnus~X-1 (located $\sim$7$^\circ$ away from the Cygnus~X region, \citealt{reid:2011}), ON\,1 (\citealt{rygl:2010a,nagayama:2011}), and G\,59.7+01 (\citealt{xu:2009a}).

There are two explanations proposed for a single connected Cygnus~X region. First, a superbubble driven by the famous Cygnus Loop supernova remnant (e.g., \citealt{walsh:1955,cash:1980}), which was recently discarded by \citet{uyaniker:2001}, and, second, an expanding Str\"omgren sphere (\citealt{mccutcheon:1970}).
Our Cygnus~X sources (all sources except AFGL\,2591) have overlapping distance uncertainties, so we cannot  use their distances for studying the structure of the complex. However, the $UVW$ space motions of the sources projected on the Galactic plane give an impression of the dynamics of the Cygnus~X complex. We plot the resulting $UVW$ space motions toward Cygnus~X in Fig. \ref{fig:small}. Apart from a clearly different behavior of AFGL\,2591, we find that W\,75N and DR\,21 have a dominant motion toward the NGP, while DR\,20 and IRAS\,20290 are moving toward the NGP and the Galactic center. These proper motions of the Cygnus~X SFRs do not suggest a common expansion center, which an expanding Str\"omgren sphere should have, so more data are needed to understand the formation of the Cygnus~X region.    

Additionally, we found that the space velocity $V$, in the direction of the Galactic
rotation (after subtracting the Galactic rotation of $V=239\,\mathrm{km~s^{-1}}$),
lies between $-10.5$ and $-8.0$\,$\mathrm{km~s^{-1}}$ for all sources
except W\,75N, where $V=+3.6\,\mathrm{km~s^{-1}}$. It seems that
most of the star-forming gas is moving with the same Galactic orbital velocity,
lagging some $9\,\mathrm{km~s^{-1}}$ behind circular orbits, as
found by most parallax studies of massive SFRs (e.g., \citealt{reid:2009b}) after taking the revised Solar motion into account (\citealt{schoenrich:2010}).

\section{Summary} 
 
We measured the trigonometric parallaxes and proper motions of five
massive SFRs toward the Cygnus~X star-forming complex
using 6.7 GHz methanol and a 22 GHz water maser. We report the following
distances:                                                              
$1.30^{+0.07}_{-0.07}$\,kpc for W\,75N, $1.46^{+0.09}_{-0.08}$\,kpc
for DR\,20, $1.50^{+0.08}_{-0.07}$\,kpc for DR\,21, $1.36^{+0.12}_{-0.11}$\,kpc
for IRAS\,20290+4052, and $3.33^{+0.11}_{-0.11}$\,kpc for AFGL\,2591.   
While the distances of W\,75N, DR\,20, DR\,21, and IRAS 20290+4052 are consistent with a single distance of $1.40\pm0.08$\,kpc for the Cygnus~X complex, AFGL\,2591 is located at a much greater distance than previously assumed. 
The space velocities of
the SFRs in Cygnus~X do not suggest an expanding
Str\"omgren sphere.

\begin{acknowledgements} 
We thank Dr. S.~Bontemps for his constructive comments, which have greatly improved this manuscript.
We are grateful to the staff at JIVE, the EVN antennas, the Japanese
VERA network, the Yamaguchi telescope, and the NRAO's VLBA and VLA for carrying
out the observations and the correlation of the data.  We thank Dr. L.~Moscadelli for his maser-plotting scripts. 
K.L.J.R is funded
by an ASI fellowship under contract number I/005/07/1. A.B. was supported by a Marie Curie Outgoing International Fellowship (FP7) of the European Union (project number 275596). Financial support by the European Research Council for the ERC Advanced Grant GLOSTAR (ERC-2009-AdG, Grant Agreement no.~247078) is gratefully acknowledged.                 
\end{acknowledgements} 

\bibliographystyle{aa} 
\bibliography{/Users/kazi/tot} 
 
\end{document}